\documentclass[twocolumn,aps,pra,showpacs, amsmath,amssymb, superscriptaddress]{revtex4}

\usepackage{graphicx}
\usepackage{dcolumn}
\usepackage{bm}

\def\ba{\begin{eqnarray}}
\def\ea{\end{eqnarray}}
\def\beq{\begin{equation}}
\def\eeq{\end{equation}}

\begin{document}

\preprint{APS/123-QED}

\title{Quantum Logic between Remote Quantum Registers}

\author{N. Y. Yao}
\affiliation{Department of Physics, Harvard University, Cambridge, MA 02138, U.S.A.}
\author{Z.-X. Gong}
\affiliation{Department of Physics and MCTP, University of Michigan, Ann Arbor, MI 48109, U.S.A.}
\author{C. R. Laumann}
\affiliation{Department of Physics, Harvard University, Cambridge, MA 02138, U.S.A.}
\affiliation{ITAMP, Harvard University, Cambridge, MA 02138, U.S.A.}
\author{S. D. Bennett}
\affiliation{Department of Physics, Harvard University, Cambridge, MA 02138, U.S.A.}
\author{L.-M. Duan}
\affiliation{Department of Physics and MCTP, University of Michigan, Ann Arbor, MI 48109, U.S.A.}
\author{M. D. Lukin}
\affiliation{Department of Physics, Harvard University, Cambridge, MA 02138, U.S.A.}
\author{L. Jiang}
\affiliation{Institute for Quantum Information and Matter, California Institute of Technology, Pasadena, CA 91125, U.S.A.}
\author{A. V. Gorshkov}
\affiliation{Institute for Quantum Information and Matter, California Institute of Technology, Pasadena, CA 91125, U.S.A.}

\date{\today}
\begin{abstract}

We analyze two approaches to quantum state transfer in solid-state spin systems.  
First, we consider unpolarized spin-chains and extend previous analysis to various experimentally relevant imperfections, including quenched disorder, dynamical decoherence, and uncompensated long range coupling.
In finite-length chains, the interplay between disorder-induced localization and decoherence yields a natural optimal channel fidelity, which we calculate.
Long-range dipolar couplings induce a finite intrinsic lifetime for the  mediating eigenmode; extensive numerical simulations of dipolar chains of lengths up to $L=12$ show remarkably high fidelity despite these decay processes.  We further consider the extension of the protocol to bosonic systems of coupled oscillators.
Second, we introduce a quantum mirror based architecture for universal quantum computing which exploits \emph{all} of the  spins in the system as potential qubits. 
While this dramatically increases the number of qubits available, the composite operations required to manipulate ``dark'' spin qubits significantly raise the error threshold for robust operation.  Finally, as an example, we demonstrate that eigenmode-mediated state transfer can enable robust long-range logic between spatially separated Nitrogen-Vacancy registers in diamond; numerical simulations confirm that high fidelity gates are achievable even in the presence of moderate disorder.

\end{abstract}

\maketitle

\section{Introduction \label{sec:intro}}

The ability to perform quantum logic between remote registers has emerged as a key challenge in the quest for scalable quantum architectures \cite{Bennett93,Kimble08, Yao12, Ladd10}. Qubits, the fundamental building blocks of such an architecture are often benchmarked by their coherence times \cite{Morton08,Tyryshkin11,Ladd05}. Naturally, those qubit implementations which possess the longest coherence times also interact most weakly with their local environment, making multi-qubit quantum logic in such systems difficult \cite{Yao12,Maurer12}. As a result, there has been tremendous recent interest in quantum data buses, which enable universal gates between physically separated quantum registers \cite{Greentree04, Blinov04,Moehring07,Togan10,Twamley10,Chudzicki10,Wu09,Plenio04}. Such data buses have been proposed in systems ranging from trapped ions \cite{Riebe08, Banchi11, King98} and superconducting flux qubits \cite{Sillanpaa07, You11,Majer07}  to coupled cavity arrays \cite{Ogden08,Bose07,deMoraesNeto11} and solid-state spin chains \cite{Bose03, Yao11,Petrosyan10, Christandl04, Burgarth07, Difranco08, Kay07, Feldman10, Clark05, Gualdi08, Wojcik05, Banchi10}. Prior proposals have focused on achieving perfect state transfer using either initialized~\cite{Bose03, Feldman10, Gualdi08, Banchi10}, engineered~\cite{Christandl04, Malinovsky97, Feldman09} or dynamically controlled quantum channels~\cite{Karbach05, Cappellaro07, Zhang09, Fitzsimons06, Fitzsimons07}. 

By contrast, here, we analyze a general method for high-fidelity quantum state transfer (QST) using an infinite-temperature (unpolarized) data bus \cite{Yao12, Yao11}. Our method requires neither external modulation during state transfer, nor precisely engineered coupling strengths within the bus, making it an ideal candidate for solid-state spin-based quantum computing architectures \cite{Yao12, Cappellaro11}. We envision the long-range coherent interaction between remote qubits to be mediated by a specific collective eigenmode of the intermediate quantum data bus.  In the solid-state, such eigenmodes naturally suffer from localization effects associated with lattice imperfections and disorder \cite{Yao11,Evers08}.  Exploration of the interplay between such localization effects and intrinsic constraints set by finite coherence times, is important to assess the feasibility of proposed architectures.

Our paper is organized as follows. In Sec.\ \ref{sec:qst}, we extend the previously proposed notion of eigenmode-mediated quantum state transfer \cite{Yao11} to the  transverse field Ising model. In addition to being closely related to the actual achievable Hamiltonian of certain driven spin systems, this simple model enables an analytic description of the state transfer protocol. In Sec.\ \ref{sec:anal}, we build upon these protocols and derive analytic expressions characterizing the channel fidelity for  state transfer between remote quantum registers. Next, we generalize our method to bosonic systems (e.g.~coupled cavities and pendulum arrays) in Sec.\ \ref{sec:bosons}. In particular, we demonstrate that chains of coupled oscillators can faithfully transport quantum information even at high oscillator temperature. 

Having described eigenmode-mediated QST in both the fermionic and bosonic context, we then turn to a specific implementation within a solid-state quantum computing architecture.  In Sec.\ \ref{sec:disorder}, we analyze eigenmode-mediated quantum state transfer between remote spin-quantum-registers \cite{Childress06, Balasubramanian09, Rittweger09}. To be specific, we consider Nitrogen-Vacancy (NV) defect center registers and examine the optimization of state transfer fidelities in the presence of both disorder and a finite  depolarization ($T_1$) time. The interplay between disorder-induced localization and decoherence yields a natural optimal channel fidelity, which we calculate.
Ultimately, this optimization demonstrates the feasibility of scalable architectures whose remote logic gates can harbor infidelities below the threshold for error correcting codes \cite{Fowler09,Wang10b}. While eigenmode-mediated QST fundamentally requires the register-chain coupling to be weaker than the intra-chain coupling, we demonstrate in Sec.\ \ref{sec:strong}, that generalizations to the strong coupling regime are also possible. We provide numerical simulations in parallel with the analytic channel fidelities derived in Sec.\ \ref{sec:anal}. 

In Sec.\ \ref{sec:lr}, we perform exact diagonalization for spin systems, which includes the full long-range  dipolar interaction. We find remarkably high fidelities for our proposed QST protocols in chains of length up to $L=12$. Finally, in Sec.\ \ref{sec:mirror}, we describe and analyze an alternate architecture, which utilizes globally controlled pulses for state transfer  \cite{Fitzsimons06,Fitzsimons07}. In this case, we demonstrate that \emph{all} spins in the system (e.g.\ even dark intermediate chain spins) can be viewed as potential qubits.  However, while this dramatically increases the number of qubits available, the composite operations required to manipulate such intermediary spin qubits significantly raise the error threshold for robust operation.


\section{Eigenmode-mediated QST \label{sec:qst}}

\noindent In this section, we begin with an idealized system in which to understand eigenmode-mediated QST \cite{Yao11}, namely, the transverse field Ising model,
\begin{equation}
H = - \sum_{i =1 }^{N-1} \kappa \sigma_{i}^{x}\sigma_{i+1}^{x} + \sum_{i=1}^{N} B \sigma_{i}^{z}
\end{equation} 
where $\kappa$ is the nearest-neighbor coupling strength and $B$ represents a uniform transverse field on each site. In addition to being realizable  in a variety of experimental systems, ranging from NVs and trapped ions to electrons floating on helium \cite{Yao11,Kim11,Mostame08}, this model also has the virtue of being exactly solvable; this will allow us to clearly illustrate the essence of eigenmode-mediated state transfer and to understand the many-body entanglement which arises. 

Expanding $\sigma_{i}^{x}$ as a function of spin flip operators, $\sigma_{i}^{\pm} = (\sigma_{i}^{x} \pm i \sigma_i^{y})/2$, and utilizing the Jordan-Wigner transformation \cite{Jordan28}, $c_i^{\dagger} = \sigma_i^{+} e^{-i \pi \sum_{j=1}^{i-1} \sigma_j^{+} \sigma_j^{-}}$, yields the fermionized Hamiltonian,
\begin{eqnarray}
H_{JW} &=& - \sum_{i=1}^{N-1} \kappa ( c_{i}^{\dagger}c_{i+1} + c_{i}^{\dagger}c_{i+1}^{\dagger} - c_{i} c_{i+1}^{\dagger} - c_{i}c_{i+1} ) \nonumber \\
&+& \sum_{i=1}^{N} B (c_{i}^{\dagger}c_{i} - c_{i}c_{i}^{\dagger})
\end{eqnarray}
\noindent which is quadratic and conserves fermionic parity without conserving particle number.  To solve $H_{JW}$, we re-express it as $\vec{\phi} \hspace{0.5mm} ^{\dagger} A \vec{\phi}$, where we define $\vec{\phi} = (c_{1}, c_{2},... ,c_{N}, c_{1}^{\dagger}, c_{2}^{\dagger},..., c_N^{\dagger})^{T}$. The matrix $A$ is real, symmetric 
and is diagonalized to 
\begin{equation}
\Lambda=
 \begin{pmatrix}
  \epsilon_1 & 0 & 0 & 0 & \cdots  \\
  0 & -\epsilon_1 & 0 & 0 & \cdots  \\
  0 & 0 & \epsilon_2 & 0 &\cdots  \\
  0 & 0 & 0 & -\epsilon_2 &\cdots  \\
  \vdots  & \vdots  & \vdots & \vdots & \ddots \\ 
 \end{pmatrix} 
\end{equation} 
via an orthogonal matrix, $O$, such that $O A O^{T} = \Lambda$. The eigenmodes come in pairs with energy $\pm \epsilon_k$, corresponding to eigenvectors $d_k = O_{2k-1,j} \phi_j$ and $d_k^{\dagger} =  O_{2k,j} \phi_j$, where $k= 1, \cdots ,N$. This transformation yields 
\begin{equation}
H_{JW} = \sum_{k=1}^{N} \epsilon_{k} (d_{k}^{\dagger} d_{k} - d_{k} d_{k}^{\dagger} ),
\end{equation} 
\noindent where the $d$-modes satisfy standard Dirac anticommutation relations. For a uniform chain the spectrum is,  $\epsilon_{k} \approx \sqrt{\kappa^{2} + B^{2} - 2B\kappa \cos q_{k}}$, where $q_{k} = k\pi/ (N+1)$.  

\begin{figure}
\includegraphics[width=3.4in]{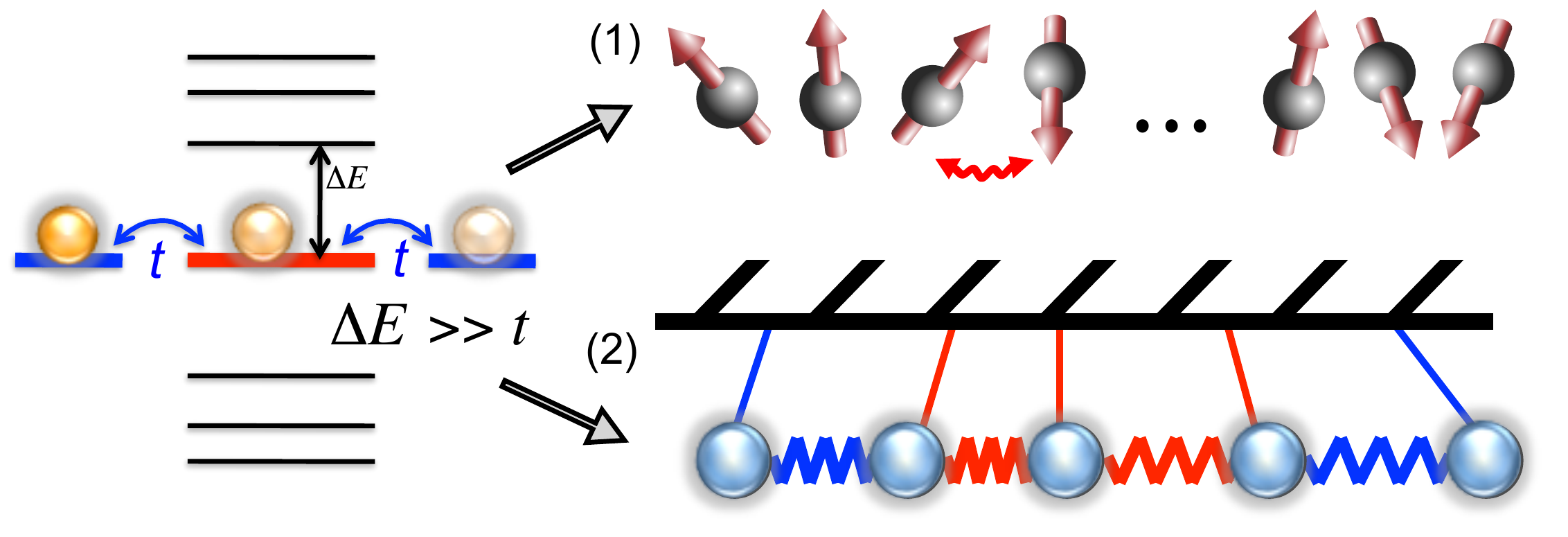}
\caption{\label{fig:Fig1} (color online). Quantum state transfer is achieved by tuning the left and right quantum registers (blue) to a particular eigenmode (red) of the intermediate data bus. By ensuring that the coupling, $t$ between the registers and the chain is sufficiently weak relative to the spacing of adjacent eigenmodes, it is possible to consider evolution in an effective three-mode picture. Such eigenmode-mediated QST is applicable in a variety of contexts, ranging from solid-state spin chains to coupled bosonic degrees of freedom (e.g. pendulums or cavity arrays).  }
\end{figure}

We now consider the addition of quantum registers, labeled $0$ and $N+1$, at the ends of the data bus (Fig.\ \ref{fig:Fig1}). The registers couple perturbatively with strength $g$ to the ends of the Ising spin chain \cite{Yao11,Feldman10} and we apply a local Zeeman field $B'$,
\begin{equation}
H' = - g ( \sigma_{0}^{x}\sigma_{1}^{x} + \sigma_{N}^{x}\sigma_{N+1}^{x} ) + B' (\sigma_{0}^{z} + \sigma_{N+1}^{z} ).
\end{equation} 
\noindent Upon fermionizing,
\begin{eqnarray}
H'_{JW} &=& -  g( c_{0}^{\dagger}c_{1} + c_{0}^{\dagger}c_{1}^{\dagger} + c_{1}^{\dagger} c_{0} - c_{0}c_{1} )  \nonumber \\
&-&  g( c_{N}^{\dagger}c_{N+1} + c_{N}^{\dagger}c_{N+1}^{\dagger} + c_{N+1}^{\dagger} c_{N} - c_{N}c_{N+1} ) \nonumber \\
&+& B' (c_{0}^{\dagger}c_{0} - c_{0}c_{0}^{\dagger} + c_{N+1}^{\dagger}c_{N+1} - c_{N+1}c_{N+1}^{\dagger}).
\end{eqnarray}
By tuning $B' = \epsilon_z$, we ensure that the external  registers are  coupled resonantly to a single finite-energy eigenmode $d_z^{\dagger}$ of the intermediate chain. Quantum state transfer proceeds via resonant tunneling through this  mode. Noting that $c_i = \sum_{k=1}^{N} (O^{T})_{i,2k-1}d_k +\sum_{k=1}^{N} (O^{T})_{i,2k}d_k^{\dagger}$ allows us to re-express $c_1$ and $c_N$ in terms of the $d$-modes.
By choosing $g O_{2z-1,1} = g O_{2z-1,N}  \ll  B', |\epsilon_z - \epsilon_{z\pm1}|$ we ensure that off-resonant eigenmodes are only weakly coupled to the quantum registers, leaving an effective three-mode picture, 
\begin{eqnarray}
H_{eff} &\approx& \epsilon_z  (d_{z}^{\dagger} d_{z} - d_{z} d_{z}^{\dagger} ) + \epsilon_z (c_{0}^{\dagger}c_{0} - c_{0}c_{0}^{\dagger}) \nonumber \\
 &+& \epsilon_z( c_{N+1}^{\dagger}c_{N+1} - c_{N+1}c_{N+1}^{\dagger}) \nonumber \\
 &-&gO_{2z-1,1} (c_{0}^{\dagger} d_{z} + d_{z}^{\dagger} c_{0} ) \nonumber \\
 &-&g O_{2z-1,N} (c_{N+1}^{\dagger} d_{z} +d_{z}^{\dagger} c_{N+1} ).
\end{eqnarray}

It is interesting to note that for $B < \kappa$, the Hamiltonian in Eq.~(2) represents a spin-less p-wave superconductor in its topological phase \cite{Kitaev01}. The zero energy boundary modes of this system have received a great deal of attention recently. As these Majorana zero modes are exponentially localized, they cannot be employed for state transfer. In our analysis, this follows from the failure of the secular approximation to remove fermion number non-conserving terms. A straight-forward calculation shows that the pairing terms precisely cancel the hopping terms in the effective evolution.

Equation (7) represents the key result of the above manipulations. State transfer is achieved by time-evolving for $\tau = \frac{ \pi}{\sqrt{2} g O_{2z-1,1}}$, leading to unitary evolution,
\begin{eqnarray}
U_{eff} &=& e^{-i \tau H_{eff}} = (-1)^{n_z} (-1)^{(c_0^{\dagger} +c_{N+1}^{\dagger})( c_0+c_{N+1})/2} \nonumber \\
&=& (-1)^{n_z} (1-(c_0^{\dagger} +c_{N+1}^{\dagger})( c_0+c_{N+1})),
\end{eqnarray}
\noindent where $n_z = d_z^{\dagger} d_z$. It is instructive to write the explicit action of $U_{eff}$ on the subspace spanned by $\Psi=\{|\Omega \rangle, c_0^{\dagger}  |\Omega \rangle,  c_{N+1}^{\dagger}  |\Omega \rangle, c_0^{\dagger} c_{N+1}^{\dagger}   |\Omega \rangle \}$, where $|\Omega \rangle$ is the vacuum associated with $c_0$, $c_{N+1}$,
\begin{equation}
 U_{eff}   \Psi = (-1)^{n_z}
 \begin{pmatrix}
  1 & 0 & 0 & 0   \\
  0 & 0 & -1 & 0  \\
  0 & -1 & 0 & 0   \\
  0 & 0 & 0 & -1  \\
 \end{pmatrix} \Psi.
\end{equation}
\noindent Up to signs,  the effective evolution in the register subspace is a swap gate. In the spin representation, owing to  Wigner strings, there exists an additional set of controlled phase (CP) gates, as shown in Fig.\ \ref{fig:qst}. Since CP$^2=\mathbb{I}$, this entanglement can be easily cancelled and logic gates between the remote registers can be successfully accomplished \cite{Yao11,Markiewicz09,Cappellaro11}. We detail two possible such protocols below. 

One protocol, herein termed ``there-and-back'', is particularly applicable to the case of multi-qubit quantum registers.  For a two-qubit register, we can label one qubit as the memory qubit while the other represents the ``coupling'' qubit. Once an eigenmode-mediated swap between the coupling qubits is accomplished, an intra-register CP-gate is then performed between the two qubits of the remote register. The return swap then cancels the unwanted entanglement illustrated in Fig.~\ref{fig:qst}, leaving only  a controlled-phase gate between the two memory qubits. Since CP gates, together with single-qubit rotations, can generate arbitrary unitary operations, such a procedure enables universal logic between remote registers.  

An alternate method, which we call the ``paired protocol'' utilizes a two-qubit encoding to cancel the Wigner strings. In this approach, the quantum information is encoded in two spins, $a$ and $b$, with logical basis $|\downarrow \rangle = | \downarrow \rangle_a |\downarrow \rangle_{b}$, $|\uparrow \rangle = | \uparrow \rangle_a |\uparrow \rangle_{b}$  \cite{Yao11,Markiewicz09,Cappellaro11}; the intuition behind this encoding is that it produces an effective bosonic excitation, thereby mitigating the effect of the fermionic Wigner strings. State transfer  proceeds by successively transferring  $a$ and $b$ through the intermediary chain.

\section{Analytic Channel Fidelity \label{sec:anal}}

We now derive the channel fidelity associated with the paired protocol. To set up the analytic  framework, we begin by calculating the fidelity of a simplified protocol, termed the ``double-swap''. In this double-swap, we consider the left register (indexed $0$) undergoing two successive  eigenmode-mediated swap gates. Ideally, this simplified protocol swaps the quantum information twice, thereby disentangling it from the intermediate chain and also returning it to its initial position at the left register. We then consider a second protocol, termed the ``single-swap'', in which the quantum information undergoes only one eigenmode-mediated swap-gate. Analyzing this protocol will illustrate the effect of the residual entanglement on the channel fidelity. Finally, we turn to the paired-protocol and demonstrate that the proposed two-qubit encoding can eliminate this entanglement, thereby enabling quantum state transfer. In Appendix \ref{sec:remotez}, we compute the channel fidelity for an eigenmode-mediated remote $\sigma^z$ gate.


\subsection{Double-swap}


\begin{figure}
\includegraphics[width=3.4in]{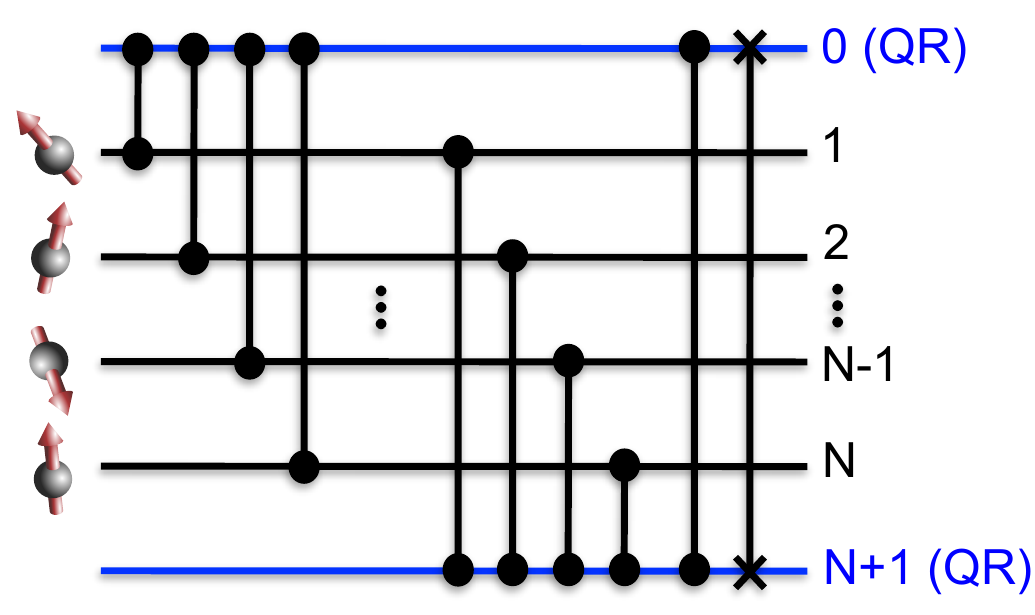}
\caption{\label{fig:qst} (color online). Schematic circuit diagram depicting eigenmode-mediated state transfer between the quantum registers (QR) $0$ and $N+1$. Controlled phase gates are represented as circle-ending dumbbells while $X$-ending dumbbells depict a swap gate. In addition to the desired state transfer, each register is CP-entangled with all intermediate spins owing to the Wigner strings associated with fermionization. This additional entanglement can be cancelled by utilizing a simple two-qubit encoding. }
\end{figure}

The average channel fidelity for a quantum dynamical operation is  given by
\begin{equation}
F =\frac{1}{2}+ \frac{1}{12} \sum_{i=x,y,z} \text{Tr} \left [ \sigma^{i} \mathcal{E} (\sigma^{i}) \right ],
\end{equation}
\noindent where $\mathcal{E}$ characterizes the quantum channel~\cite{Nielsen02}. For simplicity of notation, we will restrict ourselves to the XX-model \cite{Yao11}, $H = g( \sigma_{0}^{+}\sigma_{1}^{-} +  \sigma_{N}^{+}\sigma_{N+1}^{-}+ \text{h.c.})+\sum_{i =1 }^{N-1} \kappa( \sigma_{i}^{+}\sigma_{i+1}^{-} +\text{h.c.})$, although analogous results hold for the previously considered transverse field Ising model.  For the double-swap (DS), we let $U$ represent evolution under $H$ for a time, $t=2\tau$, equivalent to twice the state-transfer time. Let us suppose that the left register is initially disentangled from the remainder of the chain, which is in a thermal mixed state $\rho_{ch}^{DS}$; the average double-swap channel fidelity is then given by,
\begin{eqnarray}
F_{DS} &=&\frac{1}{2}+ \frac{1}{12} \sum_{i=x,y,z} \text{Tr} \left [ \sigma^{i}_0 U (\sigma^{i}_0 \otimes \rho_{ch}^{DS}) U^{\dagger}  \right ] \nonumber \\
&=&\frac{1}{2}+ \frac{1}{12} \sum_{i=x,y,z} \text{Tr} \left [ U^{\dagger}  \sigma^{i}_0 U (\sigma^{i}_0 \otimes \rho_{ch}^{DS}) \right ] \nonumber \\
&=&\frac{1}{2}+ \frac{1}{12} \sum_{i=x,y,z} \text{Tr} \left [ \sigma^{i}_0 (t) (  \sigma^{i}_0 \otimes \rho_{ch}^{DS})  \right ],
\end{eqnarray}
\noindent where $\sigma^{i}_0 (t)$ is the Heisenberg evolution of the left register.   By fermionization, this evolution can be re-expressed with respect to elements of the matrix $M =e^{-iK t}$ where $K$ is the $(N+2)\times(N+2)$ coupling matrix of the full Hamiltonian (including registers), $H = \sum_{i,j=0}^{N+1} K_{ij} c_i^{\dagger} c_j$.   Evolution of the fermi operators is governed by $\dot{c_m} = -i \sum_{n}K_{mn}c_n$, implying that $c_m(t) = \sum_{n}M_{mn}c_n$ and further, that
\begin{eqnarray}
\label{eq:sigmap}
\sigma_0^+(t) &=& U^{\dagger} \sigma^{+}_0 U = U^{\dagger} c^{\dagger}_0 U \nonumber \\
&=& \sum_i M_{0i}^{*} c_i^{\dagger} = \sum_i M_{0i}^{*} \sigma_i^{+} \prod_{l<i} e^{i \pi \sigma_l^{+} \sigma_l^{-}}, 
\end{eqnarray}
\begin{eqnarray}
\label{eq:sigmaz}
\sigma_0^z(t) &=& 2 c_0^{\dagger} (t) c_0 (t)-1 = -1 + 2\sum_{ij}M_{0i}^{*} M_{0j} c_i^{\dagger} c_j \nonumber \\
&=&-1 + 2\sum_{ij}M_{0i}^{*} M_{0j} \sigma_i^{+} \sigma_j^{-} \prod_{i<l<j} e^{i \pi \sigma_l^{+} \sigma_l^{-}}, 
\end{eqnarray}
\noindent where we have used the fact that $c_0^{\dagger}$ carries no Wigner string. To evaluate $F_{DS}$, we note that $\sigma^{\pm}=(\sigma^x\pm i \sigma^y)/2$, and hence, $\text{Tr} \left [ \sigma^{x}_0 (t) (  \sigma^{x}_0 \otimes \rho_{ch})  \right ] = \text{Tr} \left [( \sigma^{+}_0 (t) +   \sigma^{-}_0 (t))( ( \sigma^{+}_0  +   \sigma^{-}_0 ) \otimes \rho_{ch})  \right ] $.  Contributions are only obtained from the cross-terms, $\sigma^{+}_0 (t) (\sigma^{-}_0  \otimes \rho_{ch})$ and $\sigma^{-}_0 (t) (\sigma^{+}_0  \otimes \rho_{ch})$, since the number of excitations in $i=0$ must be preserved to generate a non-zero trace. For example, using Eq.~(\ref{eq:sigmap}),
\begin{eqnarray}
 &\text{Tr}& \left [ \sigma^{+}_0 (t) (  \sigma^{-}_0 \otimes \rho_{ch})  \right ] \nonumber \\
 &=&  \text{Tr} \left [ ( \sum_i M_{0i}^{*} \sigma_i^{+} \prod_{l<i} e^{i \pi \sigma_l^{+} \sigma_l^{-}} )  (  \sigma^{-}_0 \otimes \rho_{ch})  \right ] \nonumber \\
 &=&  \text{Tr} \left [ M_{00}^{*} \sigma^{+}_0  \sigma^{-}_0 \otimes \rho_{ch} \right ] = M_{00}^{*}.
\end{eqnarray}
\noindent An analogous calculation yields $\text{Tr} \left [ \sigma^{-}_0 (t) (  \sigma^{+}_0 \otimes \rho_{ch})  \right ] = M_{00}$. Finally, for the $\sigma^z$ terms, one finds, using Eq.~(\ref{eq:sigmaz}),
\begin{eqnarray}
 &\text{Tr}& \left [ \sigma^{z}_0 (t) (  \sigma^{z}_0 \otimes \rho_{ch})  \right ] = \text{Tr} \left [ -   \sigma^{z}_0 \otimes \rho_{ch}\right ]  \nonumber \\
 &+&  \text{Tr} \left [ (  2\sum_{ij}M_{0i}^{*} M_{0j} \sigma_i^{+} \sigma_j^{-} \prod_{i<l<j} e^{i \pi \sigma_l^{+} \sigma_l^{-}}  )  (  \sigma^{z}_0 \otimes \rho_{ch})  \right ] \nonumber \\
 &=&  \text{Tr} \left [ 2 M_{00}^{*} M_{00} \sigma^{+}_0  \sigma^{-}_0  \sigma^{z}_0 \otimes \rho_{ch} \right ] = 2 |M_{00}|^2,
\end{eqnarray}
\noindent where we've noted that $i=j$ to ensure that the number of excitations in each mode is conserved. Moreover, we must also have that $i=j=0$, since $\text{Tr} [\sigma^z_0]=0$. Combining the above terms yields the double-swap channel fidelity as,
\begin{equation}
F_{DS} =\frac{1}{2}+ \frac{1}{6} (M_{00} +M_{00}^{*} +|M_{00}|^2).
\end{equation}
Interestingly, we need to compute only a single matrix element to obtain the relevant channel fidelity.

\subsection{Single-swap}

We now consider the single-swap (SS) channel fidelity associated with the transfer of quantum information from the right register (indexed $N+1$) to the left register (indexed $0$),
\begin{equation}
F_{SS} =\frac{1}{2}+ \frac{1}{12} \sum_{i=x,y,z} \text{Tr} \left [ \sigma^{i}_0 (t) (\rho_{ch}^{SS} \otimes  \sigma^{i}_{N+1})  \right ],
\end{equation}
\noindent where $\rho_{ch}^{SS}$ now characterizes the initial state for spins $\{0, \cdots , N\}$. Note that $F_{SS}$ will be independent of the direction of state transfer, and we have chosen right to left for notational simplicity. From Eq.~(\ref{eq:sigmap}), one finds,
\begin{eqnarray}
\sigma_{0}^{x}(t)&=&c_{0}^{\dagger}(t)+c_{0}(t)=\sum_{i}M_{0i}^{\ast}c_{i}^{\dagger}+M_{0i}c_{i} \nonumber \\
&=&\sum_{i}[\{\text{Re}(M_{0i})\sigma_{i}^{x}+\text{Im}(M_{0i})\sigma_{i}^{y}\}\prod_{l=0}^{i-1}(-\sigma_{l}^{z})].
\end{eqnarray}
\noindent In analogy to the DS case, $i \neq N+1$ terms do not contribute to the trace,
\begin{equation}
\text{Tr}[\sigma_{0}^{x}(t)(\rho_{ch}\otimes\sigma_{N+1}^{x})]=2\text{Re}(M_{0,N+1})\text{Tr}[\rho_{ch}^{SS}\prod_{l=0}^{N}(-\sigma_{l}^{z})].
\end{equation}
The $\sigma^y$ term yields an identical contribution while the $\sigma^z$ term yields, $\text{Tr}[\sigma_{0}^{z}(t)(\rho_{ch}^{SS}\otimes\sigma_{N+1}^{z})]=2|M_{0,N+1}|^{2}$. Therefore,
\begin{equation}
\label{SSfid}
F_{SS}=\frac{1}{2}+\frac{1}{6}[2\text{Re}(M_{0,N+1})\text{Tr}[\rho_{ch}^{SS}\prod_{l=0}^{N}(-\sigma_{l}^{z})]+|M_{0,N+1}|^{2}).
\end{equation}
For perfect transfer with $F_{SS} =1$, we would require both $|M_{0,N+1}| = 1$ and $|\text{Tr}[\rho_{ch}^{SS}\prod_{l=0}^{N}(-\sigma_{l}^{z})]|=1$. In the case of an unpolarized chain, the second condition is unsatisfied since  the
expectation value of the chain parity operator $P=\prod_{l=0}^{N}(-\sigma_{l}^{z})$
is zero. The dependence of the single-swap fidelity on the intermediate chain's parity demonstrates the entanglement illustrated in Fig.~\ref{fig:qst}, and presents an obvious problem for QST. 

\subsection{Paired-Protocol}

To overcome this problem, we now turn to the two-qubit encoding proposed in Sec.~\ref{sec:qst}, i.e.\ $|\downarrow \rangle = | \downarrow \rangle_a |\downarrow \rangle_{b}$, $|\uparrow \rangle = | \uparrow \rangle_a |\uparrow \rangle_{b}$. Let us index the full chain as $\{ 0_a, 0_{b} , 1 , \cdots, N, (N+1)_{b}, (N+1)_{a} \}$ and define $U_{b}$ as the transfer process through the sub-chain $\{ 0_{b} , 1 , \cdots, N, (N+1)_{b} \}$, while $U_{a}$ represents the transfer process through the sub-chain $\{ 0_{a} , 1 , \cdots, N, (N+1)_{a} \}$. To model a realistic experimental situation, we will assume that the quantum information is originally encoded in qubit $0_a$, while qubit $0_b$ is prepared in state $|\uparrow\rangle$. A C$_{0_a}$NOT$_{0_b}$ gate is then applied to encode the information in the logical $0$-register. After the state transfer, we apply C$_{(N+1)_{b}}$NOT$_{(N+1)_a}$ to decode our quantum information into qubit $(N+1)_b$. The unitary characterizing the encoding, state transfer, and decoding is then $U =  $C$_{(N+1)_{b}}$NOT$_{(N+1)_a} U_b U_a $C$_{0_a}$NOT$_{0_b}$, and the average channel fidelity is given by
\ba
F_{enc} &=&\frac{1}{2}+ \frac{1}{12}\sum_{i=x,y,z} \text{Tr} \Big[ \sigma^{i}_{(N+1)_b} (t) (\sigma^{i}_{0_a} \otimes |\uparrow\rangle_{0_b} \langle \uparrow| \otimes \nonumber \\
&& \quad \quad \quad \quad  \quad \quad \quad \quad \otimes \rho_{ch}^{PP} \otimes \rho_{N+1}) \Big].
\ea
\noindent Here, $\rho_{ch}^{PP}$ is the mixed initial state of the intermediate chain ($\{1,\cdots,N\}$), 
while  $\rho_{N+1}$ is the mixed state of the encoded $(N+1)$ register within the logical subspace. Working within this logical subspace is crucial to ensure that CP$_{0_a,N+1_a}$CP$_{0_b,N+1_b} = \mathbb{I}$. Inspection reveals that the paired-protocol includes two contributions from the chain parity operator, and since $P^2 = \mathbb{I}$, we have effectively disentangled from the intermediate chain.  Since a consistent ordering of the spin-chain is required to implement the Jordan-Wigner transformation, the Hamiltonian, $H_{U_a}$ governing the $U_a$ transfer evolution will contain uncanceled Wigner strings. For example, the piece of $H_{U_a}$ containing the coupling between the registers and the ends of the spin-chain takes the form, $H_{U_a} = g (c_{0_a}^{\dagger} e^{i \pi n_{0_{b}}} c_1 + c_N^{\dagger}  e^{i \pi n_{(N+1)_{b}}}  c_{(N+1)_a} +\text{h.c.} )$. While one must take care to correctly evaluate such strings, an otherwise straightforward computation yields,
\begin{eqnarray}
F_{enc}&=& \frac{1}{6} ( 2|M_{0,N+1}|^2 \text{Re} \left [ M_{0,N+1}^2 - M_{0,0} M_{N+1,N+1} \right ] \nonumber \\
&+&|M_{0,N+1}|^2 + | \sum_{i=1}^N M_{N+1,i} M_{i,0} |^2) + \frac{1}{2}. \label{eq:fenc}
\end{eqnarray}
\noindent Again, one only needs to compute certain matrix elements of $M$, and, in fact, an analytic form for all such elements can be obtained (see Ref.\  \cite{Feldman10} and Appendix \ref{sec:perturb}).

Before concluding this section, we point out that one can alternatively decode the quantum information into qubit $(N+1)_a$ via C$_{(N+1)_{a}}$NOT$_{(N+1)_b}$. In this case, the expression for $F_{enc}$ is identical to Eq.\ (\ref{eq:fenc}), except the term  $| \sum_{i=1}^N M_{N+1,i} M_{i,0} |^2$ is not present. Thus, the decoding into qubit $(N+1)_b$ described above gives a slightly higher average fidelity and we will use this decoding in later numerical simulations.

\section{Generalization to Oscillator Systems \label{sec:bosons}}



\noindent In this section, we analyze the generalization of eigenmode-mediated state transfer to systems  of bosonic oscillators. The realization of such coupled-oscillators is currently being explored in systems such as, cavity arrays  \cite{Ogden08,Bose07,deMoraesNeto11}, nano-mechanical oscillators \cite{Shim07,Brown10},  Josephson junctions \cite{Cataliotti01,Teitel83,Makhlin01},  and optomechanical crystals \cite{Safavi-Naeini11}. 

Consider a chain of coupled harmonic oscillators with  Hamiltonian
\begin{equation}
H_B = \sum_{i=1}^{N} \omega a_{i}^{\dagger} a_{i} + \sum_{i=1}^{N-1} \kappa (a_{i}^{\dagger}a_{i+1} + a_{i+1}^{\dagger} a_{i}).
\end{equation}
As before, we begin by diagonalizing the Hamiltonian. Let us define $b_{k} = \frac{1}{A} \sum_j \sin \frac{jk\pi}{N+1} a_{j}$, with $A=\sqrt{(N+1)/2}$ and $k=1,\cdots,N$, yielding $H=\sum_k (\omega + \epsilon_{k})b_{k}^{\dagger}b_{k}$, where $\epsilon_{k} = 2\kappa \cos(\frac{k\pi}{N+1})$. The perturbative coupling of the two additional quantum registers at the ends of the oscillator chain is given by, $H'_B = g(a_{0}^{\dagger}a_{1} + a_{N}^{\dagger}a_{N+1} + \text{h.c.})+ \omega' (a_{0}^{\dagger}a_{0} + a_{N+1}^{\dagger}a_{N+1})$, where $g$ characterizes the register-oscillator-chain coupling strength and $\omega'$ is the register frequency. Upon re-expressing $a_1$ and $a_N$ as a function of the eigenmodes $b_k$, we arrive at the full Hamiltonian,
\begin{eqnarray}
&H_B&+\hspace{2mm}H'_B = \sum_{k=1}^{N} t_{k} (a_{0}^{\dagger}b_{k} + (-1)^{k-1}a_{N+1}^{\dagger}b_{k} +\text{h.c.})  \nonumber \\
&+&\omega' (a_{0}^{\dagger}a_{0} + a_{N+1}^{\dagger}a_{N+1})+\sum_{k=1}^{N} (\omega + \epsilon_{k})b_{k}^{\dagger}b_{k},
\end{eqnarray}
\noindent where  we let $t_{k} = (g/A) \sin [k\pi / (N+1)]$. In analogy to Sec.~\ref{sec:qst}, we consider resonant tunneling through a particular mode $b_{z}$, by tuning $\omega' = \omega+\epsilon_{z}$ and ensuring that $t_{z} \ll |\epsilon_{z}-\epsilon_{z\pm1}|$. The resulting effective Hamiltonian is $H_{eff}^B = \sqrt{2} t_{z}(\eta_{0}^{\dagger} b_{z} + b_{z}^{\dagger} \eta_{0})$, where $\eta_{0} = 1/\sqrt{2}(a_{0} +a_{N+1})$. To demonstrate state transfer, we  introduce operators $ \xi_{\pm}=1/\sqrt{2}(\eta_{0} \pm b_{z})$, yielding
\begin{equation}
H_{eff}^B = \sqrt{2}t_{z} (\xi_{+}^{\dagger}\xi_{+} + \xi_{-}^{\dagger}\xi_{-}).
\end{equation}
\noindent Let us now consider unitary evolution under $H_{eff}^B$ for a time $\tau_B = \pi/ (\sqrt{2} t_{z})$, wherein $U_{eff}^B = e^{-iH_{eff}^B\tau_B} = (-1)^{\xi_{+}^{\dagger}\xi_{+}}(-1)^{\xi_{-}^{\dagger}\xi_{-}}$, so that  $(U^B_{eff})^{\dagger} \xi_{\pm} (U^B_{eff}) = -\xi_{\pm}$. Returning to the original basis and evaluating the time evolution of $a_{0}$ and $a_{N+1}$ yields 
\begin{eqnarray}
a_{0} (\tau) \rightarrow (U^B_{eff})^{\dagger} a_{0} (U^B_{eff}) = -a_{N+1}, \nonumber \\
a_{N+1} (\tau) \rightarrow (U^B_{eff})^{\dagger} a_{N+1} (U^B_{eff}) = -a_{0},
\end{eqnarray}
\noindent demonstrating a swap gate between the oscillator-registers at the ends of the chain. As before, this state transfer is achieved \emph{independent} of the state of the intermediate chain. Moreover, there exists no additional entanglement between the registers and the intermediary oscillators; this is a direct consequence of the bosonic nature of the modes, which, unlike their Wigner-fermionic counterparts in Sec.~\ref{sec:qst},  carry no  strings.


One crucial difference with the spin-chain case is that  the occupation of the bosonic eigenmodes is
not limited to $0$ or $1$. In a highly excited system, this induces a ``bosonic enhancement'' of
off-resonant errors and will limit the achievable state transfer fidelity
as a function of temperature. In particular, the state transfer
unitary evolution gives $a_{N+1}(\tau) = M_{N+1,0} a_0 +
\sqrt{\epsilon} a_\epsilon$, where $\epsilon = 1-|M_{N+1,0}|^2 \propto g^2$
is a small error and  $a_\epsilon$ is a normalized linear combination of
the $a_i$ modes ($i = 1, \dots, N+1$). The total number of
excitations in mode $N+1$ after the state transfer is $\langle
n_{N+1}(\tau)\rangle = (1- \epsilon) \langle n_0 \rangle + \epsilon
\langle n_\epsilon \rangle$, where $n_i =  a^\dagger_i a_i$.  Therefore, if
the chain is thermally occupied with $\langle n_\epsilon \rangle \approx k
T/\omega > 1$, the coupling strength $g$ must be reduced to $g \sqrt{\omega/(k T)}$ in order to
keep errors comparable with the zero-temperature bosonic case. In realistic experimental systems, this implies an interplay between temperature, which sets the bose-enhancement of off-resonant errors and decoherence rates, which limit the minimal speed of state transfer. 

\begin{figure}
\includegraphics[width=3.4in]{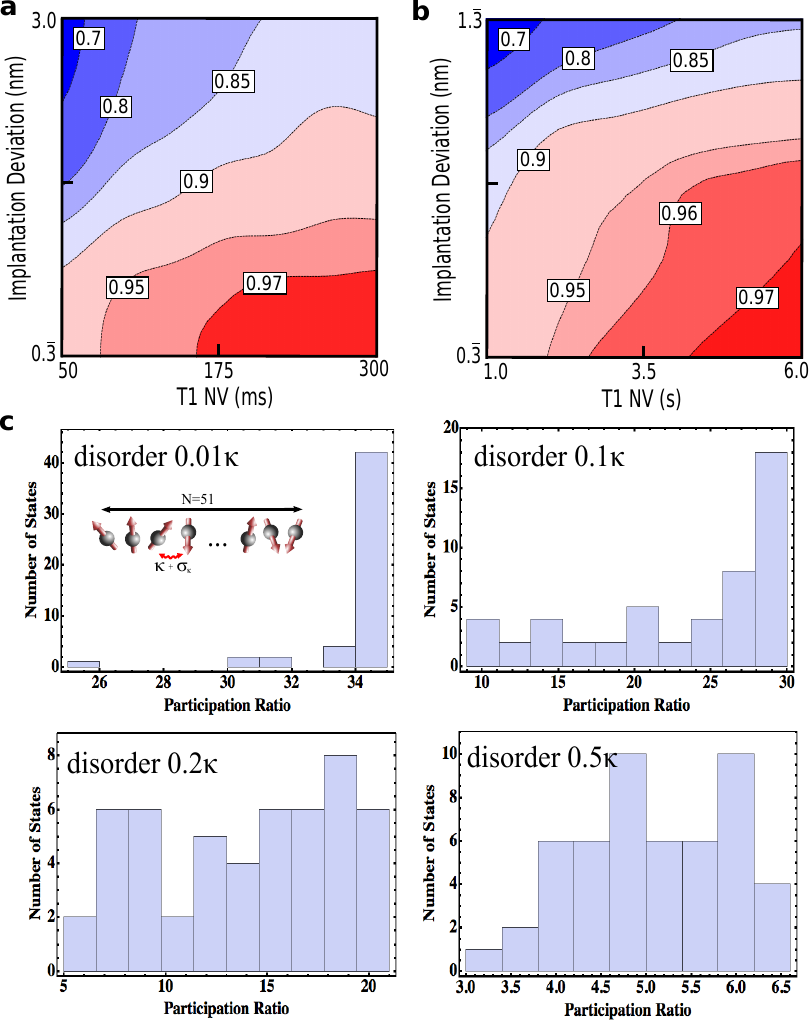}
\caption{\label{fig:avgfid} (color online). (a) Contour plots for $N=11$ characterizing the average achievable fidelity as a function of the NV depolarization time ($T_1$) and the coupling strength disorder induced by imperfect implantation. Numerics utilize an average intrachain spin spacing of $d=10$nm corresponding to a $\kappa=50$kHz dipole-dipole interaction strength. This intrachain spacing is assumed to be independent Gaussian distributed and the implantation deviation represents the standard deviation, $\sigma_d$. For each $\sigma_d$, 1000 realizations were averaged to obtain the plotted fidelity and a smooth contour plot is generated via a third-order spline interpolation. The register-chain coupling strengths $g_L$, $g_R$, Eq.~(\ref{gL}), are assumed to be fully tunable via control of the 3-level NV ground state manifold \cite{Yao11}. (b) Analogous contour plots for $N=51$. In this case, the NV registers are separated by order optical wavelength enabling individual laser manipulation without the need for subwavelength techniques. (c)  Participation ratio for eigenmodes ($N=51$). Each eigenmode is indexed by its PR and the number of states within a certain PR bin is shown. For each disorder (which are represented as fractions of the bare coupling strength $\kappa = 50$kHz), 1000 realizations are averaged.  }
\end{figure}

\section{Disorder and Decoherence \label{sec:disorder}}

Eigenmode-mediated state transfer naturally finds use in a variety of quantum computing architectures where data buses are required to connect high-fidelity remote registers \cite{Ladd10, Yao12, Cappellaro11}. Within such  architectures, it is crucial to consider an interplay between naturally occurring disorder and finite decoherence rates. While disorder in 1D  systems generically localizes all eigenmodes, leading to an exponentially long state-transfer time, in finite-size systems with weak disorder, the localization lengths can be large relative to the inter-register separation. In these cases, one must still reduce the register-chain coupling strength $g$ to compensate disorder effects, but so long as the register decay time is sufficiently long, it remains possible to achieve high-fidelity QST. In this section, we will discuss the impact of coupling-strength disorder on spin chains and will analyze the optimization of  $g$ as a function of disorder strength and qubit depolarization time.

In particular, we will consider two sources of error: 1) off-resonant coupling to alternate eigenmodes (which becomes enhanced as disorder increases) and 2) a finite register depolarization time $T_1$, 
\begin{equation}
\epsilon = \sum_{k \neq z}( g_L^2 \frac{ |\psi_{k,L}|^2}{\Delta_k^2} +g_R^2 \frac{ |\psi_{k,R}|^2}{\Delta_k^2} ) + N\frac{t}{T_1},
\end{equation}
\noindent where $g_{L(R)}$ is left (right) register-chain coupling, $\psi_{k,L(R)}$ is the eigenmode amplitude at the left (right) register, $\Delta_k$ is the energy difference from mode $z$ to mode $k$, $N$ is the chain length, $t$ is the state transfer time and $T_1$ is the depolarization time of the register. The additional factor of $N$ in the final term results from the entanglement discussed in Sec II; indeed, since each register is CP-entangled with all intermediate spins, any spin-flip of the intermediate chain immediately dephases the quantum information. 

To ensure that the tunneling rates at each end of the intermediate chain are equivalent, we envision tuning $g_L$ and $g_R$ independently, such that $t_z =g_L |\psi_{z,L}| = g_R |\psi_{z,R}|$. Plugging in for the state transfer time, $t = \pi / \sqrt{2} t_z$ yields,
\begin{equation}
\label{eq:try}
\epsilon = \sum_{k \neq z} g_L^2 \left(  \frac{ |\psi_{k,L}|^2}{\Delta_k^2} +\frac{|\psi_{z,L}|^2}{|\psi_{z,R}|^2}  \frac{ |\psi_{k,R}|^2}{\Delta_k^2} \right ) + \frac{N \pi}{\sqrt{2}T_1 g_L |\psi_{z,L}|},
\end{equation}
\noindent which enables us to derive the optimal coupling strength,
\begin{equation}
\label{gL}
g_L =  \sqrt[3]{\frac{N \pi}{2\sqrt{2}T_1  |\psi_{z,L}|} \left   ( \sum_{k \neq z}  \frac{ |\psi_{k,L}|^2}{\Delta_k^2} +\frac{|\psi_{z,L}|^2}{|\psi_{z,R}|^2}  \frac{ |\psi_{k,R}|^2}{\Delta_k^2} \right ) ^{-1}}.
\end{equation}

\subsubsection{Disorder Numerics for a Specific NV-based Architecture}

We now consider an example implementation of eigenmode-mediated state transfer in the context of a quantum computing architecture based upon Nitrogen-Vacancy (NV) registers in diamond \cite{Childress06, Balasubramanian09, Rittweger09}. Each fully controllable NV register consists of a coupled electronic and nuclear spin. The nuclear spin, with extremely long multi-second room-temperature coherence times is often thought of as the memory qubit \cite{Maurer12}, while the electronic spin, which can be optically initialized and read out, mediates interactions with other NVs \cite{Yao12,Cappellaro11}.  Our analysis of disorder effects will be based upon the specific architecture proposed in \cite{Yao12}; there, NV registers are connected by a dark-spin-chain data bus composed of spin-$1/2$ electronic spins associated with Nitrogen impurities.  One of the crucial advantages of utilizing spin chains to connect remote NVs is that this enables optical addressing of individual registers in parallel, a necessary requirement for scalable fault-tolerant quantum computation.

\begin{figure}
\includegraphics[width=3.4in]{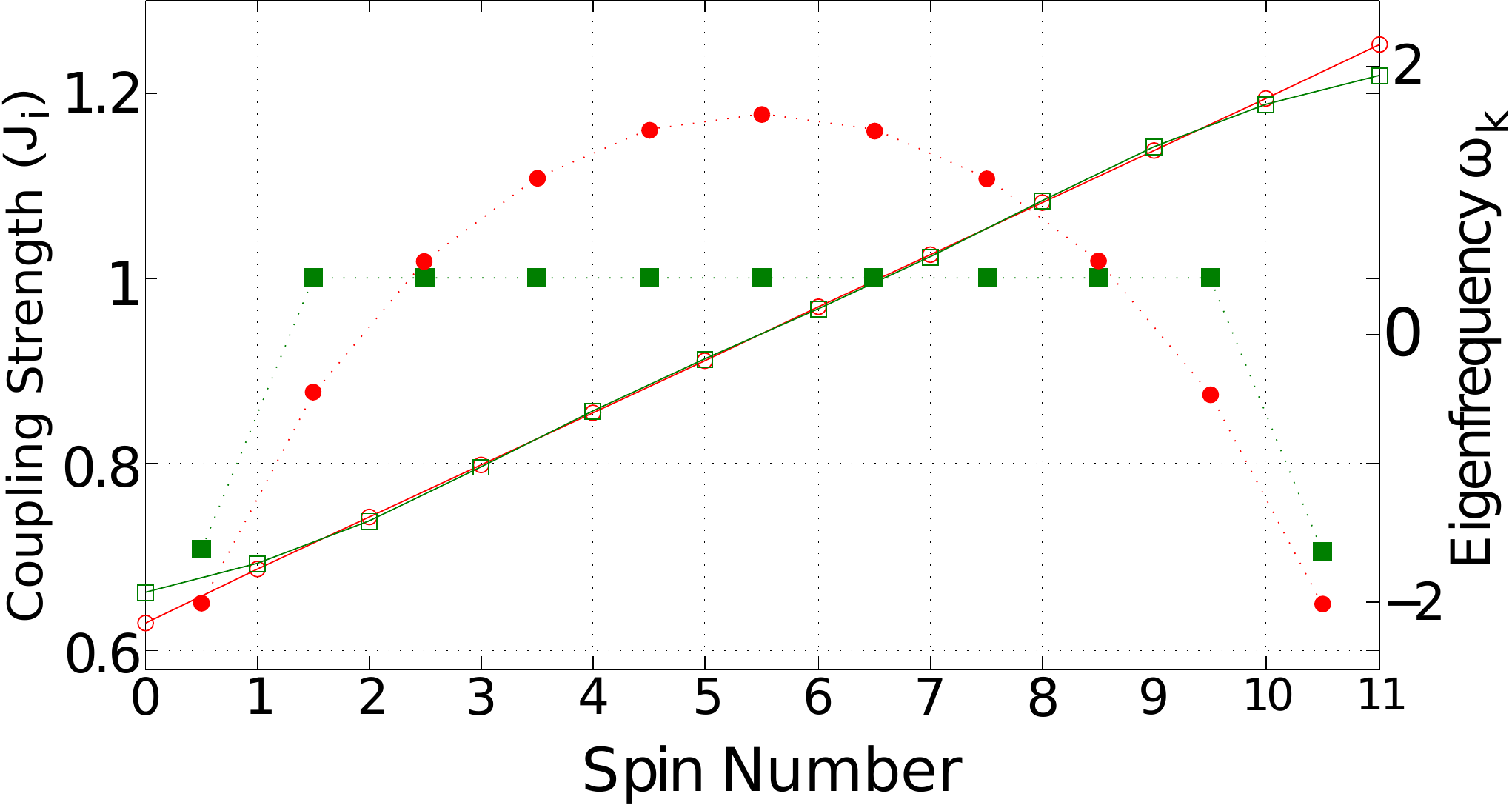}
\caption{\label{fig:quasilin} (color online). Coupling pattern $\{J_{i}\}$ between spins for two differing cases: 1) engineered couplings (circles) as in \cite{Christandl04} and 2) strong coupling regime (squares). The left y-axis characterizes the coupling strength for each case and is associated with solid symbols; the couplings are plotted between spin numbers (e.g. $J_0$ is plotted between spin number $0$ and $1$).  The right y-axis characterizes the fermionic spectrum (in this case, the x-axis is simply an index) and is associated with the open symbols. The open red circles depict the exactly linear spectrum of engineered chain, while the open green squares depict the quasi-linear spectrum of the strong coupling case with uniform interchain couplings $\kappa =1$ and optimized $g \approx 0.7$.}
\end{figure}

We consider realistic experimental parameters, with an average spin spacing of about $10$nm, corresponding to a dipole coupling strength $\approx 50$kHz. At room-temperature, NV centers are characterized by $T_1 \sim 10$ms \cite{Maurer12}, owing to an Orbach spin-lattice relaxation process; the exponential dependence of the Orbach process on temperature suggests that slight cooling can significantly extend $T_1$, with many  seconds already demonstrated at liquid Nitrogen temperatures \cite{Yao12,Jarmola11}. We now perform disorder-averaged numerics for two separate chain lengths: 1) sub-wavelength addressable ($N=11$) and 2) optical-wavelength addressable ($N=51$) \cite{Maurer10}.   We characterize the amount of disorder  by the standard deviation associated with imperfect spin positioning; in the case of NVs, the origin of this imperfection is straggle during the ion-implantation process \cite{Spinicelli11,Toyli10}. We average over 1000 disorder realizations and calculate the fidelity, $1-\epsilon$, according to  Eq.~(\ref{eq:try}); in particular, for each realization, we calculate the error for each eigenmode of the spin-chain and choose the maximum achievable fidelity. As shown in Fig.~\ref{fig:avgfid}a, high-fidelity quantum gates can be achieved for few nanometer straggle provided that the NV depolarization time is $\sim 200$ms; similarly, for the longer chain case (Fig.~\ref{fig:avgfid}b) with $N=51$, high-fidelity gates are also possible, but require significantly longer $T_1$ of a few seconds. 

Next, we analyze the  participation ratio (PR),
\begin{equation}
N_{PR} = \frac{1}{ \sum_{i=1}^{N} |\psi_i|^4}
\end{equation}
\noindent which provides a characterization of the number of sites which participate in a given eigenmode; modes are typically said to be extended if $N_{PR} \sim \mathcal{O}(N)$ and localized if $N_{PR} \ll N$. As the disorder increases, $N_{PR}$ drops sharply as depicted in the histograms in Fig.\ \ref{fig:avgfid}c. Moreover, by $\sigma_\kappa \approx 0.5 \kappa$, on average, all eigenmodes exhibit a state transfer fidelity $<2/3$ even for extremely long $T_1 \sim 5$s.

\section{Strong Register Coupling \label{sec:strong}}

\begin{figure}
\includegraphics[width=3.4in]{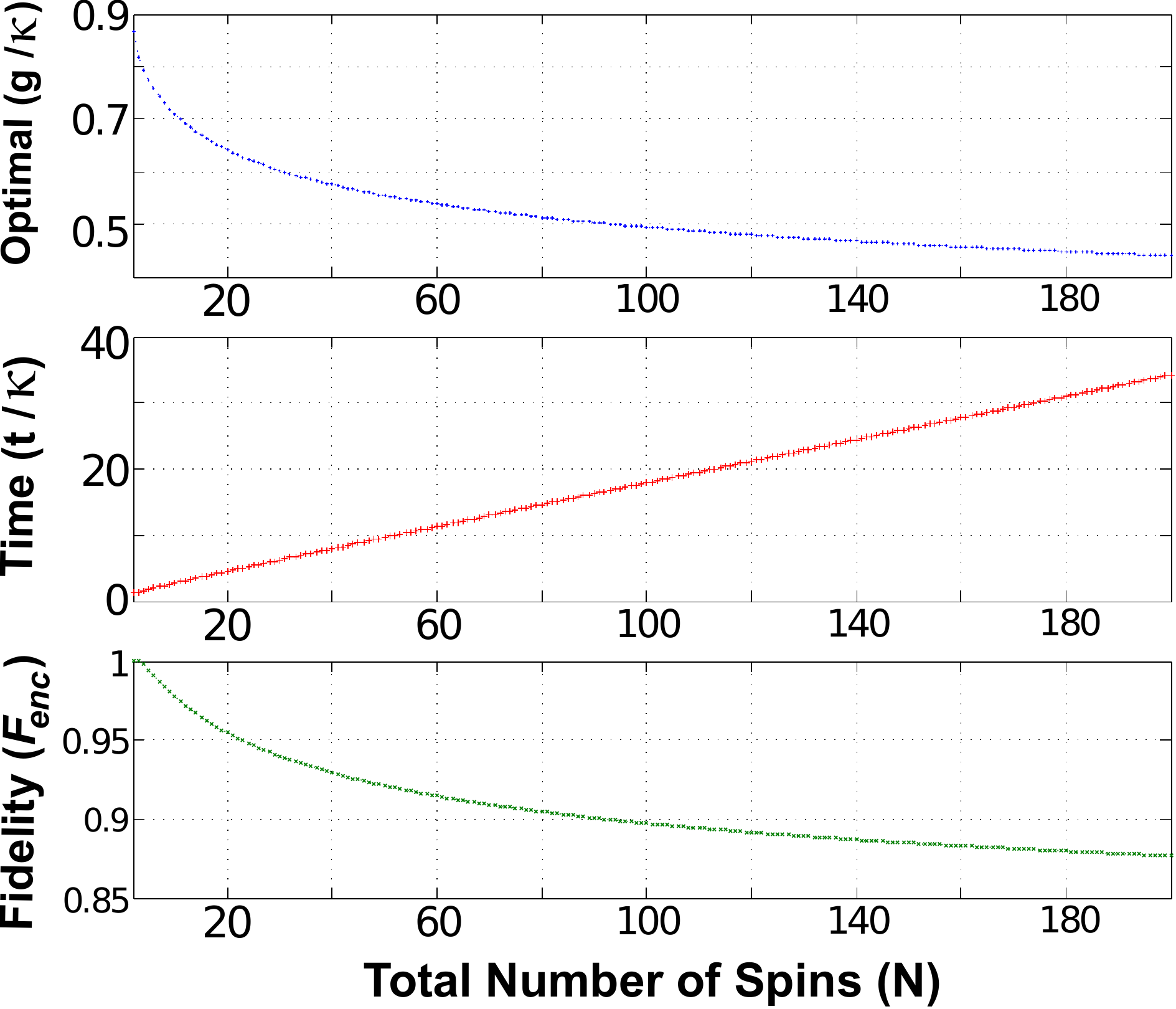}
\caption{\label{fig:scaling} (color online). Strong coupling regime: By tuning $g/\kappa \sim N^{-1/6}$,
we obtain high-fidelity QST utilizing an unpolarized chain with two-qubit
encoding (paired-protocol). The transfer time scales linearly
with $N$ (Lieb-Robinson bound) \cite{Lieb72} and high fidelities $> 90\%$ can be maintained for chain lengths up to $N=100$.}
\end{figure}

The eigenmode-mediated QST discussed above operates in the
 weak coupling regime, $g\psi \ll\kappa/N$. Numerical simulations
reveal that by optimally tuning $g=g_{M}(N) \sim \kappa$, high-fidelity QST can also be achieved (see~Fig.~\ref{fig:scaling}).
This ``strong-coupling'' regime enables faster state transfer and has been discussed in several recent studies \cite{Banchi11,Bayat11,Zwick12,Banchi10}, which focus on the case of an initially polarized intermediate chain. Here, we will demonstrate
that chains with infinite spin-temperature can nevertheless support QST in the strong-coupling regime.

To provide intuition for this strong-coupling regime, we will begin by considering the engineered
spin-chain described in \cite{Christandl04}, where we have $N+2$
spin-1/2 atoms with  nearest-neighbor XX-interactions.  The intra-chain coupling is non-uniform and is given by,  $J_{i}=\frac{1}{2}\sqrt{(i+1)(N+1-i)}$, yielding a Hamiltonian
\begin{equation}
H=\sum_{i=0}^{N}J_{i}(\sigma_{i}^{+}\sigma_{i+1}^{-}+h.c.)+\sum_{i=0}^{N+1}\frac{h}{2}\sigma_{i}^{z},
\end{equation}
where $h$ is a uniform background magnetic field.  Upon employing the Jordan Wigner transformation, we once again return to a simple tight-binding form, with $H=\sum_{ij}K_{ij}c_{i}^{\dagger}c_{j}$ where $K_{ij}=J_{i}\delta_{j,i+1} +J_{j} \delta_{i,j+1} +h\delta_{i,j}$ up to a  constant. Diagonalizing reveals $H=\sum_{k=0}^{N+1}\omega_{k}f_{k}^{\dagger}f_{k}$ with a linear spectrum given by $\omega_{k}=k+h-\frac{N+1}{2}$.


As described in Sec.~\ref{sec:anal}, the system's evolution is  governed by $c_{i}(t)=\sum_{j}M_{ij}(t)c_{i}(0)$. Upon setting $h=\frac{N+1}{2}$, one finds that at time $t=2\pi$, $M(2\pi) = \mathbb{Id}$ and therefore $c_{i}(2\pi)=c_{i}(0)$, returning the system to its initial state. As the coupling pattern $\{J_{i}\}$ harbors mirror symmetry with $J_{i}=J_{N-i}$, the orthogonal transformation, $\psi$, which diagonalizes $H$ can also be chosen mirror symmetric, $\psi_{ik}=(-1)^{N+1+k}\psi_{N+1-i,k}$. Setting $h=\frac{3}{2}(N+1)$ and $t=\pi$ yields,
\begin{equation}
M_{ij}=\sum_{k}\psi_{N+1-i,k}\psi_{jk}=\delta_{N+1-i,j}\label{eq:Mij}.
\end{equation}
To demonstrate state transfer, let us recall the analytic single-swap fidelity given by Eq.~(\ref{SSfid}). For the moment, let us assume that the spins $\{0,1,\dots N\}$ are all polarized, so  that Tr$[\rho_{ch}^{SS}P]=1$. Combined with Eq.~(\ref{eq:Mij}), which ensures $M_{0,N+1}=1$, we find  $F_{SS}=1$,
enabling perfect QST.  We note that in lieu of applying a uniform magnetic field $h=\frac{3}{2}(N+1)$, one
can also just apply a simple phase gate $U_{P}=\begin{pmatrix}1 & 0\\
0 & (-i)^{N+1}
\end{pmatrix}$ on spin $0$ following transfer. 

Turning now to the case of an unpolarized spin chain, we again employ the two-qubit encoding previously described. 
In this case, one will need to apply the phase gate, $U_{P}^{2}=\begin{pmatrix}1 & 0\\
0 & (-1)^{N+1}
\end{pmatrix}$ to the logical qubit after state transfer. 

The state transfer fidelities for these two strong coupling methods are given analogously by,
\begin{eqnarray}
F_{SS} & = & \frac{1}{2}+\frac{1}{6}[2|M_{0,N+1}|+|M_{0,N+1}|^{2}),\\
F_{enc} & = & \frac{1}{2}+\frac{1}{6}[2|M_{0,N+1}|^{2}|M_{0,N+1}^{2}-M_{0,0}M_{N+1,N+1}|\nonumber \\
 &  & +|M_{0,N+1}|^{2}+|\sum_{i = 1}^N M_{N+1,i}M_{i,0}|^{2}].
\end{eqnarray}
While these expressions are valid for an arbitrary coupling pattern (so long as the resultant fermionic Hamiltonian is quadratic), to ensure high-fidelity QST, we require $|M_{0,N+1}|\approx1$.  As depicted in Eq.~(\ref{eq:Mij}), satisfying this constraint is intimately related to the linear spectrum resulting from the choice of $J_{i}=\frac{1}{2}\sqrt{(i+1)(N+1-i)}$. 

Let us now consider the strong coupling regime ($g \sim \kappa$) where $J_{0}=J_{N}=g$ and $J_{1}=J_{2}=...=J_{N-1}=\kappa$. Surprisingly, tuning only $g/\kappa$ enables one to obtain a quasi-linear spectrum \cite{Banchi11}; such a spectrum will then ensure that $|M_{0,N+1}|\approx1$, as desired. Of course, for $N=2,3$, $J_{i}=\frac{1}{2}\sqrt{(i+1)(N+1-i)}$
can be  satisfied exactly. Although for $N>3$, an exactly linear spectrum cannot be obtained, it is possible to optimally tune $g = g_M(N)$, so that $\omega_{k}$ looks nearly identical to the previous linear spectrum, $k-\frac{N+1}{2}$ ($h=0$), as shown in Fig.\ \ref{fig:quasilin}.  In particular, by optimizing $F_{enc}$, we obtain $g_{M}\sim N^{-1/6}$, with a transfer time $\tau\sim N$ (Fig. \ref{fig:scaling}), consistent with \cite{Banchi11}.




\section{Long-range Interactions \label{sec:lr}}

\begin{figure}
\includegraphics[width=3.4in]{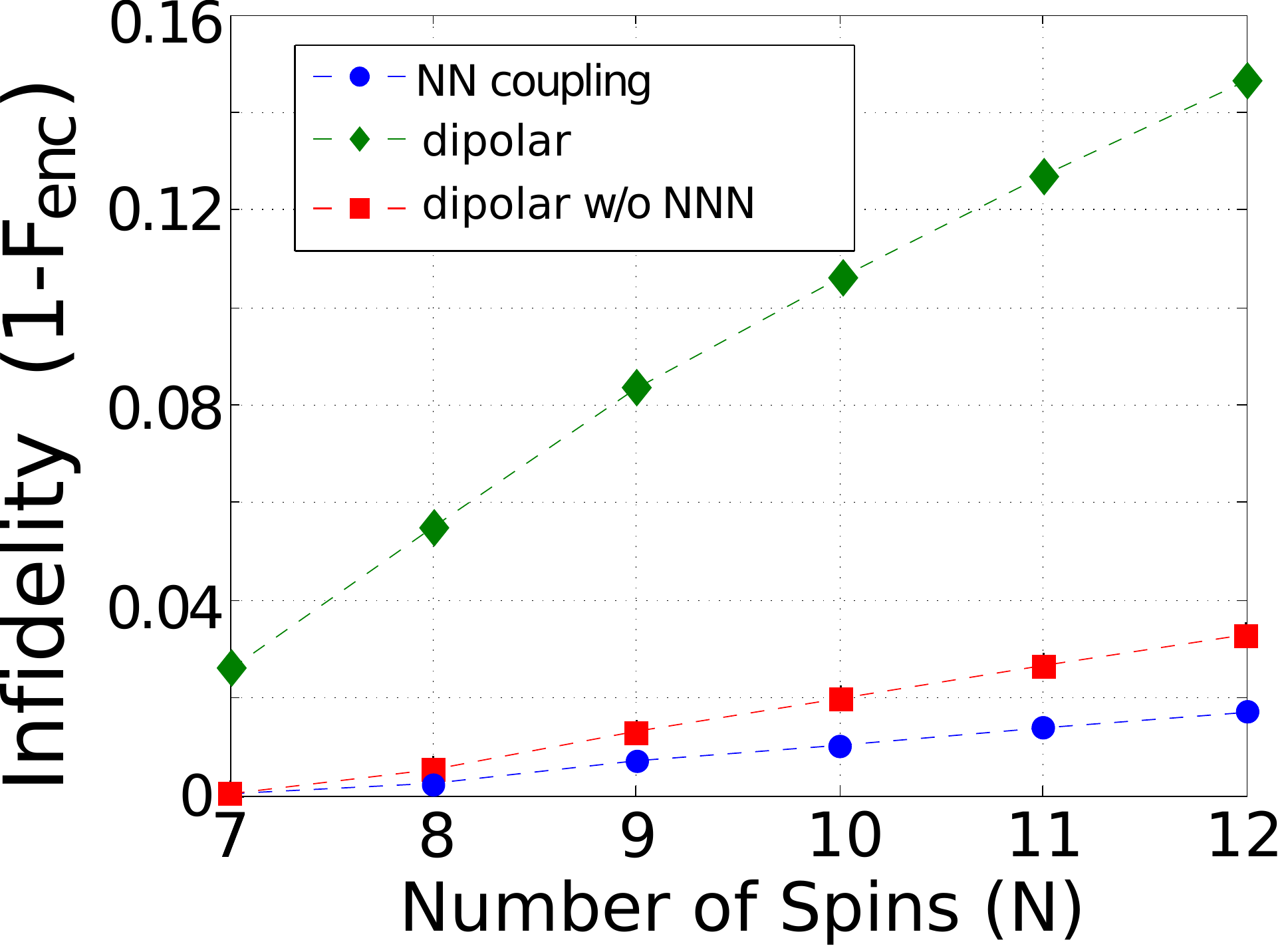}
\caption{\label{fig:stronginfid} (color online). Infidelity of QST for a strongly coupled chain with long range interactions ($1/T_1$ is assumed negligible).
 Encoded state transfer (paired protocol) fidelities are shown for dipolar (diamonds), NNN-canceled-dipolar (squares) and NN interaction (circles) models.   }
\end{figure}

Finally, we now consider the effect of longer range interactions.  The majority of proposals for spin-chain state transfer focus on  approximate nearest-neighbor models; however, the microscopic magnetic dipolar interaction is naturally long-range and decays as $1/r^3$, inducing an important infidelity in quantum state transfer.  The origin of this infidelity becomes especially evident as we examine the Jordan-Wigner fermionization of the spin chain. Each Wigner fermion carries a string of the form $e^{-i \pi \sum_{j=1}^{i-1} \sigma_j^{+} \sigma_j^{-}}$. In the nearest-neighbor case, all such strings cancel pairwise leaving a simple quadratic model. However, with longer-range interactions, uncanceled strings remain and generically introduce perturbative quartic terms into the Hamiltonian. These quartic terms imply that the model, unlike the transverse field Ising model,  is no longer diagonalizable in terms of free fermions. In the previous free fermion case, the energy of each eigenmode is independent of the occupation of all other eigenmodes; this enables state transfer even when the spin-temperature of the chain is effectively infinite.  By contrast, the quartic terms associated with the long-range dipolar coupling introduce interactions between fermionic eigenmodes; the energy fluctuations of each eigenmode, caused by changing occupations of other modes, naturally dephases  quantum information, limiting the operational spin temperature of the chain. 

Certain proposals have suggested the possibility of using dynamical decoupling to effectively cancel next-to-neareast neighbor (NNN) interactions \cite{Yao12}, but the complete canceling of all long-range interactions requires a level of quantum control that is currently beyond the realm of experimental accessibility. 
Since any long-range $XX$ coupling destroys the quadratic nature of the fermionic Hamiltonian, an analytic solution for state transfer fidelities in the presence of full dipolar interactions is not available. Thus, we perform exact diagonalization for chains of length up to $N=12$ (total number of spins), as shown in Fig.~\ref{fig:stronginfid}. We obtain the encoded state transfer fidelities for dipolar, NNN-canceled-dipolar and NN interaction models. Remarkably, even with full dipolar interactions, fidelities $\sim 90\%$ can be obtained for a total of $N=10$ spins; in the case where NNN interactions are dynamically decoupled, the fidelities can be further improved to $\sim 98\%$ at similar lengths.

\section{Quantum Mirror Architecture \label{sec:mirror}}

 In this section, we present an alternate quantum computing architecture based upon pulsed quantum mirrors \cite{Fitzsimons06,Fitzsimons07}. By contrast to eigenmode-mediated state transfer, remote quantum logic will be achieved by global rotations and NN Ising interactions. To remain consistent, we choose to discuss the advantages and disadvantages of such an architecture within the context of NV registers. In particular, analogous to Sec.~\ref{sec:disorder}, we consider NV registers connected by spin $1/2$ chains of implanted  Nitrogen impurities.

\begin{figure}
\includegraphics[width=3.4in]{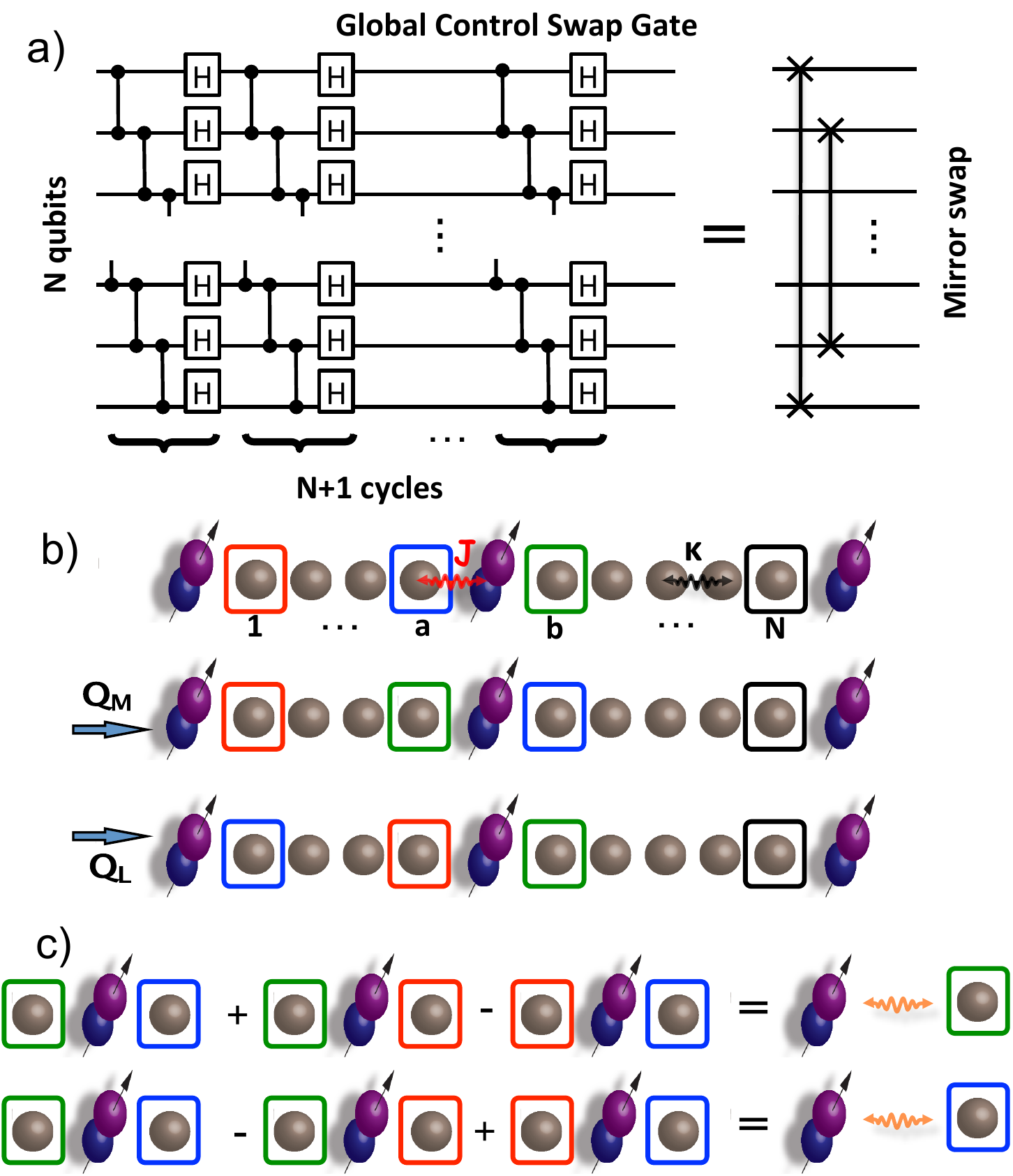}
\caption{\label{fig:Fig7} (color online). (a) In a qubit chain of length $N$, a mirror swap operation is defined as the pairwise swap between the $(1,N)$, $(2,N-1)$, $\cdots$ qubits. This can be achieved via global control in the form of single qubit gates (Hadamards) and controlled phase gates. Regardless of the initial state, a mirror swap occurs after a $N+1$ cycles of $Q = \tilde{H} \cdot \tilde{CP}$ \cite{Fitzsimons06}, where $\tilde{H}$ represents a global Hadamard operation and $\tilde{CP}$ denotes a global controlled phase operation. (b)  Coupling the central NV register to the NV on the left (right) requires the ability to perform a directed swap to a neighboring Nitrogen spin (grey). These directed swap operations are made possible by using combinations of the mirror sequences $Q_{M}$ (swapS a pair of impurities directly surrounding an NV) and $Q_{L}$ (mirror swaps an individual impurity chain). (c) Utilizing a fast echo-pulse on the NV register (in combination with $Q_{M}$ and $Q_{L}$)    allows one   to generate selective interactions between the NV and any outlined Nitrogen.  }
\end{figure}

Let us begin with a detailed discussion of the mixed spin system composed of NV centers and Nitrogen impurities \cite{Yao12}.  The full Hamiltonian of a single Nitrogen impurity is,
\begin{equation}
H_{N} = -\gamma_{e} \vec{B} \cdot \vec{S} - \gamma_{N} \vec{B} \cdot \vec{I} + A_{\parallel} S^{z}I^{z} + A_{\perp} (S^{x}I^{x}+S^{y}I^{y}),  
\end{equation}

\noindent where $\vec{S}$ is the spin-$1/2$ electronic spin operator, $\vec{I}$ is the nuclear spin operator, and $A_{\parallel} = -159.7$MHz, $A_{\perp} = -113.8$MHz are the hyperfine constants associated with the Jahn-Teller axis.

We envision the application of a magnetic field and field gradient, which, within a secular approximation,  reduces the Hamiltonian of a nearest neighbor Nitrogen-impurity chain to Ising form \cite{Yao12}, 
\begin{equation}
\label{HN}
H_{N} = \kappa \sum_{i=1}^{N-1}{S^{z}_{i}S^{z}_{i+1}} + \sum_{i=1}^{N}{(\omega_{0} + \delta_{i})S^{z}_{i}},
\end{equation}

\noindent where $\kappa$ is the relevant component of the dipole tensor, $\omega_{0}$ captures the electronic Zeeman energy, and $\delta_{i}$ characterizes the hyperfine term, which is nuclear-spin-dependent, for each impurity. 
Taking into account the magnetic dipole coupling between the electronic spin of the NV register and the surrounding Nitrogen impurities allows us to consider the mixed spin system,
\begin{equation}
H_{eff} =  \sum_{i=1}^{a-1}{\kappa S^{z}_{i} S^{z}_{i+1}}+ J S^{z}_{NV} (S^{z}_{a} + S^{z}_{b}) + \sum_{i=b}^{N-1}{\kappa S^{z}_{i} S^{z}_{i+1}},
\end{equation}
\noindent where $J$ is the strength of register-impurity interaction, the Zeeman term in Eq.~(\ref{HN}) is assumed to be echoed out, and superscripts $a$, $b$ represent the pair of nearest-neighbor impurities next to a given register (assuming for simplicity a 1D geometry as shown in Fig.~\ref{fig:Fig7}).  The selective individual addressing of the NV registers is accomplished via a combination of optical beams and microwave driving; this enables an isolation of the coupling between the NV register and the two neighboring impurities. In particular, it is possible to perform unitary evolution of the form $U_{eff} =  e^{-i H_{eff} T'/2} S^{x}_{NV} e^{-i H_{eff} T'/2} S^{x}_{NV}= e^{-i \kappa \sum{S^{z}_{i}S^{z}_{i+1}} T'}$ and hence,
\begin{equation}
U_{local} =  e^{-i H_{eff} T}  e^{-i \kappa \sum{S^{z}_{i}S^{z}_{i+1}} T'} = e^{-i J S^{z}_{NV} (S^{z}_{a} + S^{z}_{b}) T}
\end{equation}

\noindent by choosing $\kappa(T+T') = 2\pi m$ for integer $m$. We note that this condition implies that the fidelity of $U_{local}$ is extremely sensitive to both coupling-strength disorder as well as the general long-range nature of the dipolar interaction.

\subsection{Globally Controlled Mirror swap} 

\noindent Considering only global addressing of the Nitrogen spin-chain and unitary evolution as described above, we demonstrate a universal set of operations between remote NV registers. Coherent register coupling is achieved by means of global pulses which mirror the quantum state of the impurity chain \cite{Fitzsimons06}; the pulses take the form of Hadamard gates and controlled phase gates, which can be generated by evolution under an Ising Hamiltonian. In an impurity spin-chain of length $N$, the global pulses swap the state of the first and $N^{th}$ spin, the state of the second and $(N-1)^{st}$ spin etc, as shown in Fig.\ \ref{fig:Fig7}a. The total mirror swap  results from $N+1$ cycles of Hadamard and controlled phase gates on all impurities, $Q_{n+1} = (\prod{H_{i}} \cdot \prod{CP_{i}})^{n+1}$. This globally controlled impurity mirror will ultimately enable the directed and coherent interaction between remote NV registers.

\begin{figure}
\includegraphics[width=3.6in]{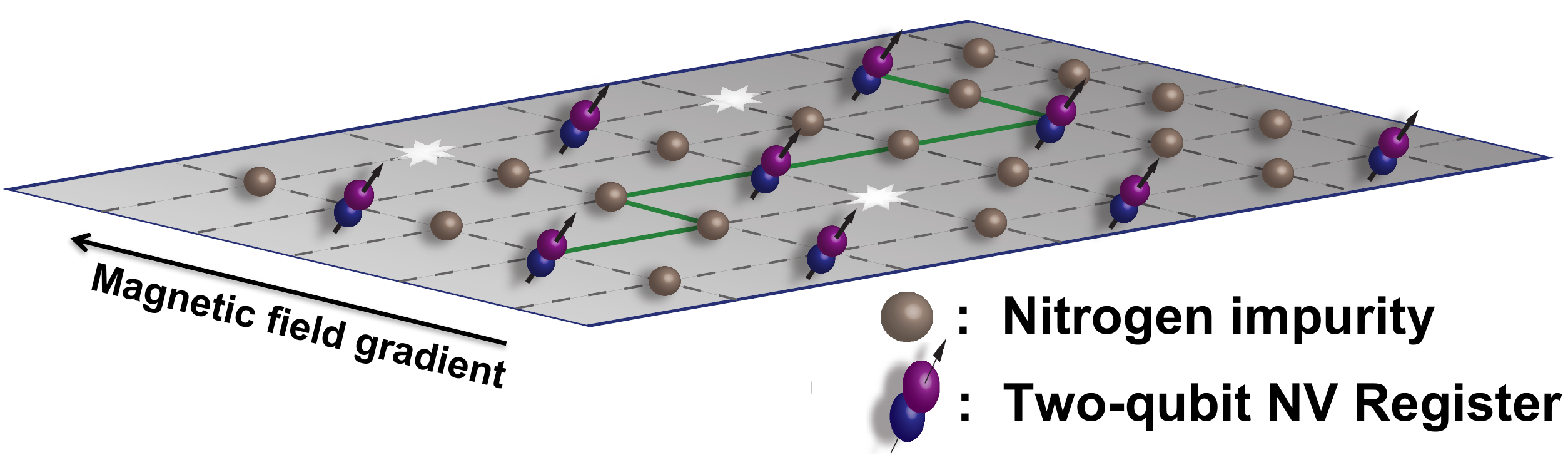}
\caption{\label{fig:Fig8} (color online). Schematic diagram of the 2D computational lattice showing limitations imposed by missing Nitrogen implantations (stars) and imperfect Nitrogen-to-NV conversions. Coherent coupling of distant NV registers in a faulty 2D array can be achieved via global pulsed control of a  spin-chain quantum data bus. A combination of optical beams and a magnetic field gradient allows for individual control of NV registers; combined with global single qubit operations on Nitrogen impurities in any given row (orthogonal to the field gradient) \cite{Yao12}, directed swap operations (e.g. dark green path) can be achieved, which allow for quantum information transfer along arbitrary paths. This field gradient enables a swap gate to be performed between two NV registers in adjacent rows, which occupy the same column.  Moreover, it in fact also enables any pair of rows to be swapped, provided that the intrarow interactions refocus.   }
\end{figure}

Let us now consider a specific NV register, separated from neighboring registers by impurity spin-chains on both sides, as shown in Fig.\ \ref{fig:Fig7}b.  Since the Ising Hamiltonian generates a controlled phase gate, it is possible to achieve a mirror swap between any set of qubits connected by   Ising interactions. In particular, the impurity Ising interaction allows for a mirror operation within any impurity chain, while the Ising interaction corresponding to $U_{local}$ allows for a three qubit mirror centered around any  NV register. This local unitary enables an operation, $Q_{M}$, which swaps the state of the Nitrogen neighbors of the central NV register as shown in Fig.\ \ref{fig:Fig7}b. 

To couple the central NV register to a specific side register, it will be necessary to break the left-right symmetry of the Ising interaction; this is achieved by exploiting the length asymmetry between Nitrogen chains to the left and right of the NV register. Indeed, it is often possible to refocus the mirror operation in one impurity chain while causing the edge impurity pair to swap in the other chain; we will denote this operation as $Q_{L}$, as shown in Fig.\ \ref{fig:Fig7}b. Combinations of $Q_{M}$ and $Q_{L}$ successfully manipulate and permute the impurities such that the nearest neighbors of the central NV register can be any pair of the three impurities (blue, red, green), as depicted in Fig.\ \ref{fig:Fig7}c.  In combination with local rotations of the central register, this enables the application of ``directed'' unitary evolution, e.g. $U_{directed} =  e^{-i JS^{z}_{NV}S^{z}_{N_{b}} T}$, allowing for the NV register to selectively couple to either side. This enables an  interaction between any pair of neighboring NV registers effectively mediated by a single Nitrogen impurity,
\begin{equation}
H_{med} = J (S^{z}_{NV_{1}} + S^{z}_{NV_{2}})  S^{z}_{N_{b}},
\end{equation}

\noindent where $NV_{1}$ and $NV_{2}$ denote the neighboring registers to be coupled and $N_{b}$ represents the mediating impurity. The form of this Ising interaction implies that an application of $Q_{M}$ on this effective three qubit system will swap the quantum information of the two electronic spins of the remote NV registers. Since each NV center harbors a nuclear-spin qubit in addition to its electronic spin \cite{Childress06}, the ``there-and-back'' protocol described in Sec.~\ref{sec:qst} enables universal logic between remote registers.

Having achieved the ability to coherently couple distant NV registers within a row, assisted by Nitrogen impurities, we now turn to the coupling between adjacent rows in a two-dimensional lattice (Fig.~\ref{fig:Fig8}). The simplest approach involves applying a magnetic field gradient along the columns. This would enable a swap gate to be performed between two NV registers in adjacent rows, which occupy the same column, provided all other interactions are echoed out. The limited occurrence of vertically adjacent NVs is a significant source of overhead; however, this limitation can be overcome if we achieve  the ability to swap any pair of nearest neighbor qubits in the two-dimensional array, essentially allowing for the construction of arbitrary paths (Fig.~\ref{fig:Fig8}). Moreover, the ability to swap along arbitrary paths also provides an elegant solution to the experimental limitation imposed by implantation holes, where a Nitrogen impurity may be missing from the ideal 2D lattice. Finally, it also enables the use of nominally dark Nitrogen impurities as computational resources, thereby significantly increasing the number of effectively usable qubits. 

\begin{figure}
\includegraphics[width=3.4in]{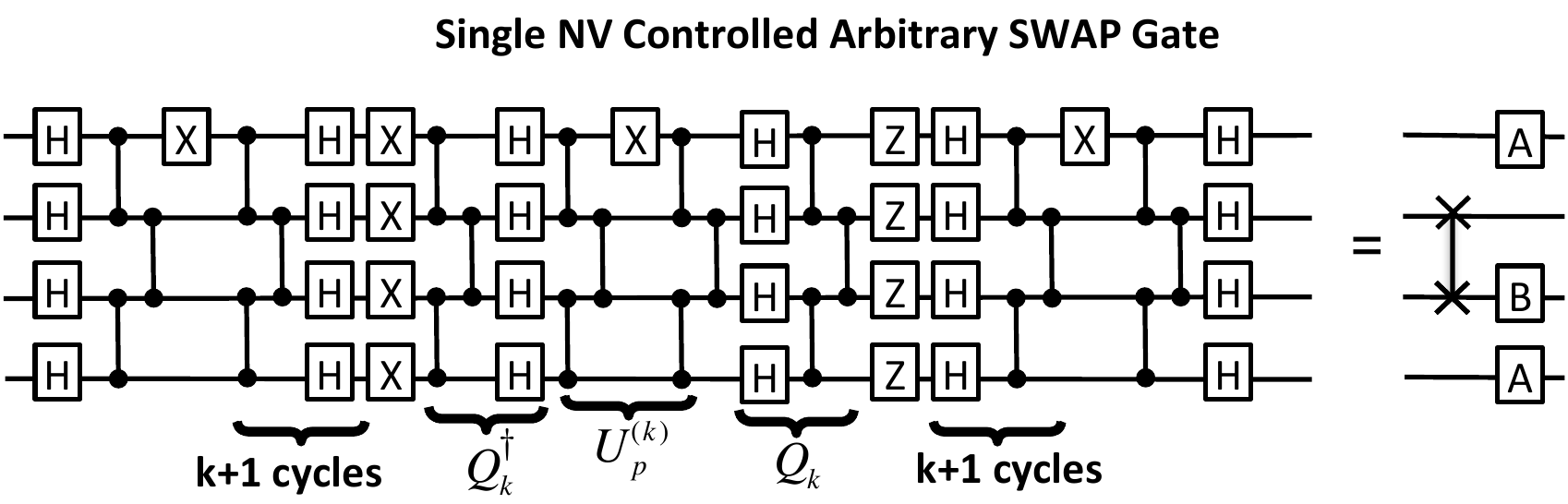}
\caption{\label{fig:arbitrary} (color online). Individual control of any given NV register (row 1) enables a swap operation between any two neighboring qubits along the same row. We illustrate the specific example of a four qubit chain. The depicted gate sequence achieves a swap gate (up to individual qubit rotations $A$, $B$) between the second and third qubit by only applying a local gate $X$ on the first qubit (NV register) and global operations elsewhere. }
\end{figure}

While arbitrary individual control of impurities would trivially enable such a scheme, realistic constraints limit us to individual control of NV registers and only global control of the impurity chains. Thus, it is necessary to utilize the permutation operation inherent to individual cycles $(\prod{H_{i}} \cdot \prod{CP_{i}})^{n+1}$ of the mirror operation. These gate cycles correspond to an effective propagation of local gates via a relabelling of qubits within a given chain. In the simplest scenario, it is possible to apply a swap gate between the second and third qubit by only utilizing local rotations on the first qubit and global operations elsewhere, as shown in Fig.\ \ref{fig:arbitrary}. The fundamental operation to be propagated is $U_{p} = \tilde{CP} \cdot X_{1} \cdot \tilde{CP}$ where $X_{1}$ is an $x$ rotation (by $\pi$) on the first qubit and $\tilde{CP}$ represents a global controlled phase gate; propagation takes the form of conjugation by mirror cycles where $Q_{k}=(\tilde{H} \cdot \tilde{CP})^{k}$ and $U_{p}^{(k)} = Q_{k}^{\dagger} U_{p} Q_{k}$. To apply a swap operation on the $n$ and $n+1$ qubit, we let $k=n-1$ and apply
\begin{equation}
U_{swap} = \tilde{H} U_{p}^{(k)} \tilde{H} \tilde{X} U_{p}^{(k)} \tilde{Z}  \tilde{H} U_{p}^{(k)}  \tilde{H},
\end{equation}
\noindent where $\tilde{X}$ is a global $x$ rotation and $\tilde{Z}$ is a global $z$ rotation (by $\pi$).  This protocol requires the ability to produce a boundary at the location of the first qubit and allows for swaps between arbitrary spins in a given row; moving quantum information between rows  can be achieved provided intrarow interactions refocus (e.g.~if vertical and horizontal nearest-neighbor distances differ).


\section{Conclusions \label{sec:conc}}

In summary, we have extended the analysis of eigenmode-mediated state transfer to a variety of imperfections ranging from disorder-driven localization to uncompensated long-range interactions. By calculating the analytic channel fidelity associated with eigenmode-mediated state transfer, we clarify the effects of entanglement arising from the protocol and illustrate the method in which the two-qubit encoding overcomes this challenge.  We analyze our protocol in the context of proposed solid-state quantum computing architectures; numerical simulations with realistic experimental parameters reveal that  QST errors can be kept below certain surface-code error-correcting thresholds.  Furthermore, we have generalized our protocol to the case of bosonic oscillator systems. This approach may enable the routing of a ``ground-state-cooled'' mode through a relatively ``hot'' intermediate oscillator chain, thereby significantly reducing the resources associated with system-wide cooling. 

Moreover, our work may also provide insight into generalized infinite-temperature state transfer. In particular, by introducing a time-dependent control of the register-chain coupling, one may be able to compensate for off-resonant errors. This approach finds analogy to the continuum wave-packet limit, where dispersion limits transfer fidelities; in this case, pre-shaping of the packet can overcome nonlinearities of the dispersion.  

Finally, we describe an alternate architecture based upon global control pulses which also enables remote quantum logic; in particular, we demonstrate that even intermediate chain spins can be used as registers, despite the fact that they are unable to be individually addressed. This may provide the blueprint for a novel quantum computing architecture which utilizes dark spins as quantum memory resources.



\section{Acknowledgements}

We gratefully acknowledge the insights of and conversations with Pierre Meystre, Alex Zhai, Sidd Viswanathan, Peter Komar and Hendrik Weimer.  This work was supported by the NSF, DOE (FG02-97ER25308), CUA, 
 DARPA, AFOSR MURI and NIST. AVG acknowledges support from the Lee A. DuBridge Foundation. LJ acknowledges support from the Sherman Fairchild Foundation and the NBRPC. AVG and LJ  acknowledge the IQIM, an NSF Physics Frontiers Center with support of the Gordon and Betty Moore Foundation. The Michigan group was supported by the ARO and the AFOSR MURI program, the IARPA MUSIQC program, the DARPA OLE program, and the NBRPC(973 Program) 2011CBA00300 (2011CBA00302).

\appendix
\section{Perturbative Calculation of Channel Fidelity \label{sec:perturb}}

As an extension of the analytic fidelity derivations presented in Sec.~\ref{sec:anal}, here, we provide a closed form expression for certain relevant matrix elements. We will work perturbatively ($g \ll \kappa/\sqrt{N}$) and will begin with the case of odd chain length. Let us consider computing $1-|\langle 0|M|N+1\rangle|^2$ for $z = (N+1)/2$. Recall that $K$ is the $(N+2) \times (N+2)$ coupling matrix of the full XX Hamiltonian. We can characterize it with basis $|j\rangle$ ($j = 0,1,...,N+1$) and express
\begin{equation}
K= g (|0\rangle \langle 1| + |N\rangle \langle N+1| + \textrm{h.c.}) + \sum_{j = 1}^{N-1} (|j\rangle \langle j+1| + \textrm{h.c.}),
\end{equation}
\noindent where we have set the intrachain coupling strength $\kappa=1$.
The time required for eigenmode-mediated state transfer is $t = \sqrt{N+1} \pi/(2 g)$ and $M = \exp(-i K t)$. Let us now define $|\pm\rangle = (|0\rangle \pm |N+1\rangle)/\sqrt{2}$ and further suppose that $N = 4x-3$ for $x\in \mathbb{Z}_{>0}$ (we will consider the other case below). After going into the diagonal basis $|k\rangle$ ($k=1,\cdots,N$) of the intermediate chain, the Hamiltonian breaks down into two decoupled parts as follows:
\begin{eqnarray}
K &=& K_+ + K_-, \\
K_- &=& \sum_{k = \textrm{even}} \left[\Delta_k |k\rangle \langle k| + \Omega_k  (|-\rangle  \langle k| + \text{h.c.})\right], \\
K_+ &=& \sum_{k = \textrm{odd}} \left[\Delta_k |k\rangle \langle k| + \Omega_k  (|-\rangle  \langle k| + \text{h.c.})\right],
\end{eqnarray}
where $\Delta_k = 2 \cos[\pi k/(N+1)]$ and $\Omega_k = (2 g/\sqrt{N+1}) \sin[\pi k/(N+1)]$. First consider $K_-$, which deals only with even $k$ and does not deal with the zero-energy mode $|z\rangle \equiv |(N+1)/2\rangle$. The eigenstates are perturbed only slightly from the original states and we call them $|\tilde -\rangle$ and $|\tilde k\rangle$ with energy $0$ and $\Delta_k + O(g^2)$, respectively. Moreover, we have 
\begin{equation}
|-\rangle \approx \left(1 - \sum_k \left(\frac{\Omega_k}{\Delta_k}\right)^2\right) |\tilde -\rangle - \sum_k \frac{\Omega_k}{\Delta_k} |\tilde k\rangle.
\end{equation}
So
\begin{equation}
\langle -| e^{-i K_- t}|-\rangle \approx 1 - 2 \sum_{k < z} \left(\frac{\Omega_k}{\Delta_k}\right)^2 (1-\cos(\Delta_k t)), \label{eq:hm}
\end{equation}
where as discussed, the sum here is only over even $k$.

We now consider $K_+$, which deals with odd $k$ and is a little more difficult to treat since it has the zero-energy mode. The eigenstates are $|\tilde s\rangle$ (for symmetric) with energy $\Omega + O(g^3)$, $|\tilde a\rangle$ (for antisymmetric) with energy $-\Omega + O(g^3)$, and $|\tilde k\rangle$ (for all odd $k$ except for $k = z$) with energy $\Delta_k + O(g^2)$. We find
\begin{equation}
|+\rangle \approx \frac{1}{\sqrt{2}} \left(1 - \sum_{k < z} \left(\frac{\Omega_k}{\Delta_k}\right)^2\right) (|\tilde s\rangle  +|\tilde a\rangle) - \sum_{k \neq z} \frac{\Omega_k}{\Delta_k} |\tilde k\rangle.
\end{equation} 
So
\begin{equation}
\langle +| e^{-i K_+ t}|+\rangle \approx -1 + 2 \sum_{k < z} \left(\frac{\Omega_k}{\Delta_k}\right)^2(1+\cos(\Delta_k t)), \label{eq:hp}
\end{equation}
where the sum  is only over odd $k$.

\begin{figure}
\includegraphics[width=3.4in]{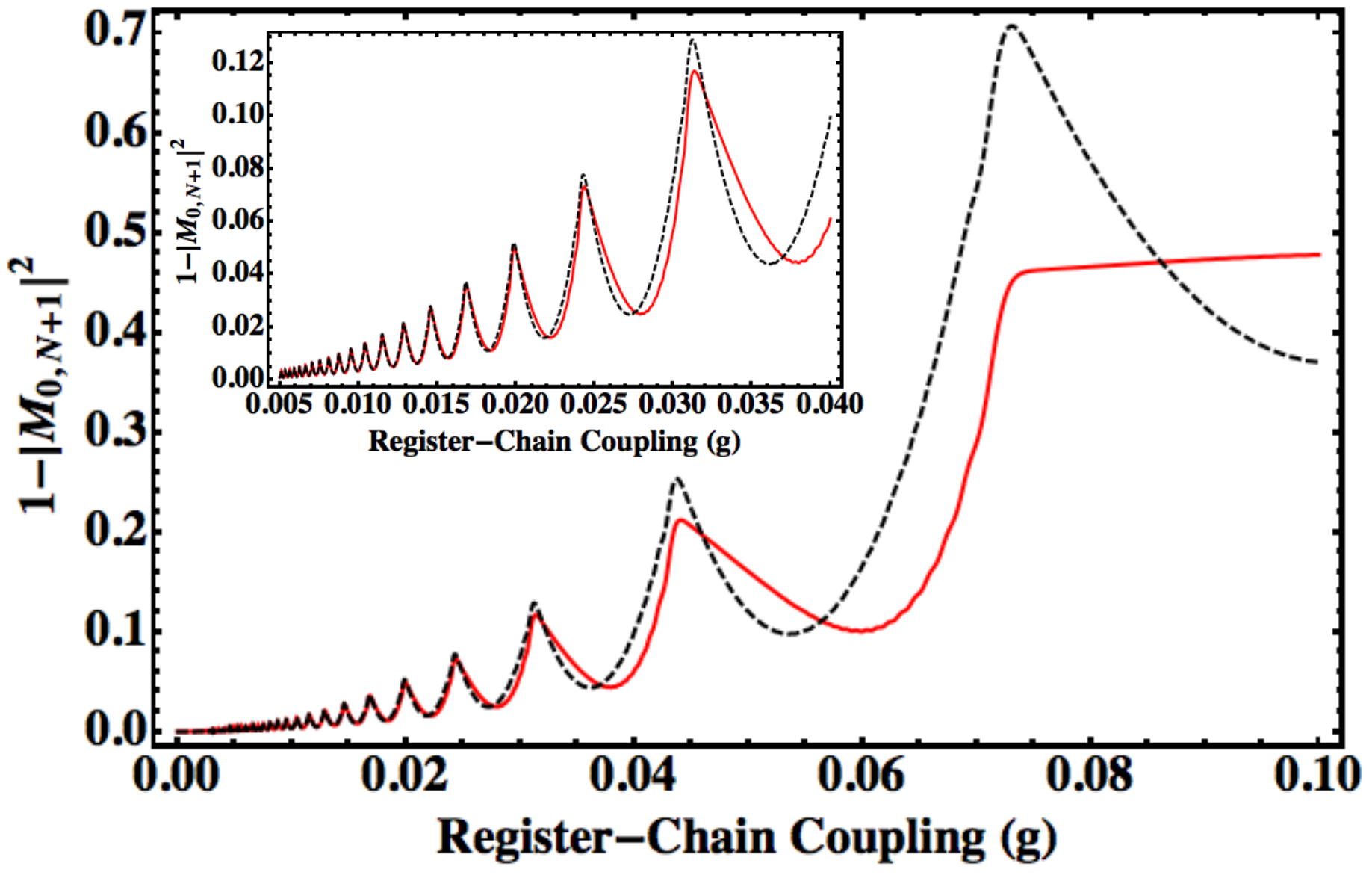}
\caption{\label{fig:perturb} (color online). Depicts a comparison between the perturbative (dotted) and exact (solid) calculation for the matrix element $M_{0,N+1}$ for $N=205$. $1-|M_{0,N+1}|^2$ is plotted as a function of $g$, the register-chain coupling strength ($\kappa =1$). 
As expected, perturbation theory breaks down as $g\sim \kappa /\sqrt{N}$. The inset depicts a zoomed in region for $0.005 < g < 0.04$, where the perturbative expression is in excellent agreement with the exact calculation. }
\end{figure}

Putting the results together, we obtain
\begin{eqnarray}
1&-& |\langle 0|M|N+1\rangle|^2  \nonumber\\ 
&=& 1 - \frac{1}{4} |\langle+|e^{-i K_+ t}|+\rangle - \langle-|e^{-i K_- t}|-\rangle|^2 \nonumber \\
&\approx & 1 - \frac{1}{4} \left( -2 + 2 \sum_{k < z} \left(\frac{\Omega_k}{\Delta_k}\right)^2(1- (-1)^k \cos(\Delta_k t)\right)^2 \nonumber \\
&\approx& 2 \sum_{k < z} \left(\frac{\Omega_k}{\Delta_k}\right)^2(1- (-1)^k \cos(\Delta_k t)),
\end{eqnarray}
where the sum  is over both odd and even $k$ less than $z \equiv (N+1)/2$.  Generalizing to all odd $N$, we find
\begin{eqnarray}
1&-&|\langle 0|M|N+1\rangle|^2 \nonumber \\
& \approx & 2 \sum_{k < z} \left(\frac{\Omega_k}{\Delta_k}\right)^2(1+ (-1)^{k+z} \cos(\Delta_k t)),
\end{eqnarray}
where $z = (N+1)/2$ and where the sum is over both odd and even $k$ (Fig.\ \ref{fig:perturb}). 

Let us now generalize our expression to include the case of even $N$.  Keeping $\Delta_k = 2 \cos[\pi k/(N+1)]$ and $\Omega_k = (2 g/\sqrt{N+1}) \sin[\pi k/(N+1)]$, we now have $z = N/2$, $t = \pi/\Omega_z$, and states $0$ and $N+1$ are shifted by $\Delta_z + \delta$ (note that $\delta \sim \mathcal{O}(g^2)$). We now define $\tilde \Delta_k = \Delta_k - \Delta_z$ and shift all energies by $\Delta_z$, so that $|0\rangle$ and $|N+1\rangle$ are now at energy $\delta$. Since we are interested in the square of the matrix element, this overall energy shift will not affect our result. We take $K_\pm$ and $|\pm \rangle$ as before. As before, we first consider $N = 4x$ for $x \in \mathbb{Z}_{>0}$ (the remaining even $N$ will be discussed below).

We calculate the evolution of $|+\rangle$, which is coupled to all odd $|k\rangle$ and, thus, not coupled to $|z\rangle$. We have
\begin{eqnarray}
e^{-i K_+ t} |+\rangle &\approx& \left(1- \frac{1}{2} \sum_k \left(\frac{\Omega_k}{\tilde \Delta_k}\right)^2\right)|\tilde + \rangle e^{i A} \nonumber\\
&-& \sum_k \frac{\Omega_k}{\tilde \Delta_k} |\tilde k\rangle e^{- i \tilde \Delta_k t},
\end{eqnarray} 
where $A = - \delta t + \sum_k \frac{\Omega_k^2}{\tilde \Delta_k} t$. To second order,
\begin{equation}
\langle +| e^{-i K_+ t}|+\rangle \approx e^{i A} - \sum_k  \left(\frac{\Omega_k}{\tilde \Delta_k}\right)^2 (1-\cos(\tilde \Delta_k t)),
\end{equation}
where the sum is over odd $k$. We now consider the evolution of $|-\rangle$ (which is coupled to all even $|k\rangle$, including $|z\rangle$). We find
\begin{eqnarray}
e^{-i K_- t}|-\rangle &\approx& \frac{1}{\sqrt{2}} \left(1- \frac{1}{2} \sum_{k \neq z} \left(\frac{\Omega_k}{\tilde \Delta_k}\right)^2\right) (|\tilde s\rangle + |\tilde a\rangle) (-1) e^{i B} \nonumber \\
 &-& \sum_{k \neq z} \frac{\Omega_k}{\tilde \Delta_k} |\tilde k\rangle e^{- i \tilde \Delta_k t},
\end{eqnarray}
where $B = \left(- \delta t + \sum_k \frac{\Omega_k^2}{\tilde \Delta_k} t\right)/2$. Thus, 
\begin{equation}
\langle -| e^{-i K_- t}|-\rangle \approx - e^{i B} + \sum_{k \neq z}  \left(\frac{\Omega_k}{\tilde \Delta_k}\right)^2 (1+\cos(\tilde \Delta_k t)),
\end{equation}
where the sum is over even $k$. Putting the calculations together, we find
\begin{eqnarray}
1&-&|\langle 0|M|N+1\rangle|^2 = 1 - \frac{1}{4} |\langle+|e^{-i K_+ t}|+\rangle - \langle-|e^{-i K_- t}|-\rangle|^2 \nonumber \\
&\approx & \sum_{k \neq z} \left(\frac{\Omega_k}{\tilde \Delta_k}\right)^2(1+ (-1)^k \cos(\tilde \Delta_k t)) + \frac{1}{4} (A-B)^2.
\end{eqnarray}
Generalizing to all even $N$, we have,
\begin{eqnarray}
1&-&|\langle 0|M|N+1\rangle|^2 \nonumber \\
&\approx & \sum_{k \neq z} \left(\frac{\Omega_k}{\tilde \Delta_k}\right)^2(1+ (-1)^{k+z} \cos(\tilde \Delta_k t)) \nonumber \\
&+& \frac{1}{4} (A-B)^2,
\end{eqnarray}
where
\begin{eqnarray}
A &=&  \frac{3 + (-1)^z}{4} \left[- \delta t + \sum_{\textrm{odd } k \neq z} \frac{\Omega_k^2}{\tilde \Delta_k} t \right]\\
B &=&  \frac{3 - (-1)^z}{4} \left[- \delta t + \sum_{\textrm{even } k \neq z} \frac{\Omega_k^2}{\tilde \Delta_k} t \right].
\end{eqnarray}
Thus, by setting
\begin{eqnarray}
\delta = \sum_{k \neq z} \frac{1-3 (-1)^{z+k}}{2} \frac{\Omega_k^2}{\tilde \Delta_k},
\end{eqnarray}
 we obtain $A - B = 0$, yielding
\begin{eqnarray}
1&-&|\langle 0|M|N+1\rangle|^2  \nonumber \\
&\approx & \sum_{k \neq z} \left(\frac{\Omega_k}{\tilde \Delta_k}\right)^2(1+ (-1)^{k+z} \cos(\tilde \Delta_k t)),
\end{eqnarray}
which holds for both even and odd $N$. One should note that tuning $\delta$ only affects small $N$, since for larger $N$,  $(A-B)^2/4$ is negligible.

We now compute $M_{0,0}$ employing the techniques outlined above; moreover, we note that any requisite matrix element entering the average channel fidelity formulae can be computed in a similar fashion. For the $M_{0,0}$ case, the $K_-$ expression remains identical to Eq.\ (\ref{eq:hm}).
However, the $K_+$ expression, Eq.\ (\ref{eq:hp}), is now different,
\begin{equation}
\langle +| e^{-i K_+ t}|+\rangle \approx 1 - 2 \sum_{k < z} \left(\frac{\Omega_k}{\Delta_k}\right)^2(1-\cos(\Delta_k t)),
\end{equation}
where the sum is over odd $k$. Combining yields,
\begin{eqnarray}
1- M_{0,0} &=& 1 - \frac{1}{2} \left[\langle+|e^{-i K_+ t}|+\rangle + \langle-|e^{-i K_- t}|-\rangle \right] \nonumber \\
&=& \sum_{k < z} \left(\frac{\Omega_k}{\Delta_k}\right)^2(1-\cos(\Delta_k t)),
\end{eqnarray}
where the sum is now over both odd and even $k$.

\section{Channel Fidelity for Remote $\sigma^z$ Gate \label{sec:remotez}}

In this appendix, we illustrate the channel fidelity associated with an eigenmode-mediated remote $\sigma^z$ gate. In combination with the detailed discussion of the double-swap channel fidelity in Sec.~\ref{sec:anal}, this provides the framework for calculating the gate fidelity of a remote controlled-phase gate. In particular, we examine the process whereby: 1) register-$0$ is swapped across the intermediate chain, 2) a $\sigma^z$ gate is performed at register-$N+1$ and 3) a second return step of eigenmode-mediated state transfer is performed. In the ideal case, this remote $\sigma^z$ channel should result in a $\sigma^z$-gate on register-$0$ and hence, the associated fidelity is given by
\begin{eqnarray}
\label{remotez}
F_z =\frac{1}{2}+ \frac{1}{12} \sum_{i=x,y,z} \text{Tr}_0 \left [ \sigma^{z}_0 \sigma^{i}_0 \sigma^{z}_0  \text{Tr}_A  \left [U_{z} (\sigma^{i}_0 \otimes \rho_{ch}) U_z^{\dagger} \right ] \right ] \hspace{2mm}
\end{eqnarray}
where $U_{z} = U \sigma^z_{N+1} U$, $U$ represents an eigenmode-mediated swap, $\rho_{ch}$ is the mixed state of spins $1,\cdots,N+1$, $\text{Tr}_0$ traces over register-$0$ and $\text{Tr}_A$ traces over all other spins. Let us begin by calculating the time evolution of $\sigma^{+}_0$,
\begin{eqnarray}
\sigma^{+}_0 (t) &=& U^{\dagger}_z c_0^{\dagger} U_z = U^{\dagger} e^{i \pi n_{N+1}}U^{\dagger} c_0^{\dagger} U e^{i \pi n_{N+1}} U \nonumber \\
&=& U^{\dagger} \left [ M_{0,0}^{*} c_0^{\dagger} - M_{0,N+1}^{*} c_{N+1}^{\dagger} + \sum_i M_{0,i}^{*} c_{i}^{\dagger} \right ] U \nonumber \\
&\rightarrow&  (M_{0,0}^{*})^2 c_0^{\dagger} - (M_{0,N+1}^{*})^2 c_{N+1}^{\dagger} + \sum_i (M_{0,i}^{*})^2 c_{i}^{\dagger}, \hspace{7mm}
\end{eqnarray}
where we have used the fact that the number of excitations in each mode must be preserved. As before, for $i=x,y$, only cross terms involving $\sigma^+$ and $\sigma^-$ provide a non-zero contribution to Eq.~(\ref{remotez}). We find
\begin{eqnarray}
 \text{Tr}_0 \left [ \sigma^{z}_0 \sigma^{+}_0 \sigma^{z}_0  \text{Tr}_A  \left [U_{z} (\sigma^{-}_0 \otimes \rho_{ch}) U_z^{\dagger} \right ] \right ] \nonumber \\
= (M_{0,N+1}^{*})^2 -  (M_{0,0}^{*})^2 - \sum_i (M_{0,i}^{*})^2.
\end{eqnarray}
\noindent An analogous calculation yields the contribution from the opposite cross term and thus, we now turn to the $\sigma^z$ contribution. Again, we begin by calculating the time evolution,
\begin{eqnarray}
\sigma^{z}_0 (t) &=&  U^{\dagger} e^{i \pi n_{N+1}}U^{\dagger} (2c_{0}^{\dagger} c_0 -1) U e^{i \pi n_{N+1}} U \nonumber \\
&\rightarrow&2 U^{\dagger} e^{i \pi n_{N+1}}  \sum_{i,j} M_{0,i}^{*}M_{0,j} c_{i}^{\dagger} c_{j} e^{i \pi n_{N+1}}U,
\end{eqnarray}
where we have dropped the $(-1)$ contribution from the first line since it will ultimately trace to zero.  Conjugation by $e^{i \pi n_{N+1}} $ affects $  \sum_{i,j} M_{0,i}^{*}M_{0,j} c_{i}^{\dagger} c_{j} $ only if $i$ or $j$ equals $N+1$; in these cases, the matrix element gets an additional negative sign. We can capture this by defining an $(N+2) \times (N+2)$ diagonal matrix $S$, which contains unity along all diagonal entries except the last, where it contains $(-1)$. Using $S$, we find,
\begin{eqnarray}
\sigma^{z}_0 (t) &=&  2U^{\dagger}  \sum_{i,j} \tilde{M}_{0,i}^{*} \tilde{M}_{0,j} c_{i}^{\dagger} c_{j} U \nonumber \\
&\rightarrow&2 \sum_{i,j} \tilde{M}_{0,i}^{*} \tilde{M}_{0,j} \sum_{i',j'} M_{i,i'}^{*} M_{j,j'} c_{i'}^{\dagger} c_{j'},
\end{eqnarray}
where $\tilde{M} = MS$. A non-zero contribution arises only if $i'=j'=0$, wherein we find $\text{Tr} [\sigma^+_0 \sigma^-_0 \sigma^z_0]=1$. Combining all contributions yields,
\begin{eqnarray}
F_z &=&\frac{1}{2}+ \frac{1}{6} \left [  (M_{0,N+1}^{*})^2 -  (M_{0,0}^{*})^2 - \sum_i (M_{0,i}^{*})^2  + \text{c.c.}\right ] \nonumber \\ 
&+& \frac{1}{6} \sum_{i,j} \tilde{M}_{0,i}^{*} \tilde{M}_{0,j} M_{i,0}^{*} M_{j,0}  \nonumber \\
&=& \frac{1}{2}+ \frac{1}{6} \left [   |\langle 0 | MSM | 0 \rangle|^2  -2 \text{Re} (\langle 0 | MSM | 0 \rangle)   \right ],
\end{eqnarray}
where we have made use of the fact that $M$ is symmetric.

\end{document}